\documentstyle[12pt,subeqnarray]{article}
\newlength{\mytopmargin}
\newlength{\myleftmargin}
\def\rlx{\relax\leavevmode}
\def\zz{\rlx\hbox{\small \sf Z\kern-.4em Z}}
\def\rr{\rlx\hbox{\scriptsize \rm I\kern-.18em R}}
\def\qq{\rlx\hbox{\,$\inbar\kern-.3em{\rm Q}$}}
\newcommand{\bin}[2]{\left( #1 \atop #2 \right)}
\newcommand{\ca}{C^{(\alpha)}}

\newcommand{\ka}{\kappa}
\newcommand{\si}{\sigma}
\newcommand{\la}{\lambda}
\newcommand{\al}{\alpha}

\begin{document}
\vspace{3.5cm}
\noindent
\begin{center}{  \Large\bf
 THE CALOGERO-SUTHERLAND MODEL AND \\[2mm] GENERALIZED 
CLASSICAL POLYNOMIALS}
 \end{center}
 \vspace{5mm}

 \noindent
 \begin{center} T.H.~Baker\footnote{email: tbaker@maths.mu.oz.au}\\
 {\it Department of Mathematics, University of Melbourne, \\
  Parkville, Victoria 3052, Australia } \\[3mm]
 P.J.~Forrester\footnote{email: matpjf@maths.mu.oz.au}
 \footnote{Permanent address:
{\it Department of Mathematics, University of Melbourne, 
 Parkville, \\\hspace*{3.7cm} Victoria 3052, Australia} }
 \\
 {\it Research Institute for Mathematical Sciences, \\Kyoto University,
 Kyoto 606, Japan}
 \end{center}
 \vspace{.5cm}

\begin{quote}
Multivariable generalizations of the classical Hermite, Laguerre
and Jacobi polynomials occur as the polynomial part of the eigenfunctions
of certain Schr\"odinger operators for Calogero-Sutherland-type quantum systems.
For the generalized Hermite and Laguerre polynomials the multidimensional
analogues of many classical results regarding generating functions,
differentiation and integration formulas, recurrence relations and summation
theorems are obtained. We use this and related theory to evaluate the global
limit of the ground state density, obtaining in the Hermite case the Wigner
semi-circle law, and to give an explicit solution for an initial value problem
in the  Hermite and Laguerre case.
\end{quote}
\date{}

\setcounter{equation}{0}
\section{Introduction}

\noindent
The Calogero-Sutherland model refers to exactly solvable quantum many body 
systems in  one-dimension with pair potentials proportional to $1/r^2$ (in 
some asymptotic limit at least),
and which have exact BDJ--type ground states:
\begin{equation}\label{gstate}
\psi_0 = \prod_{j=1}^N f_1(x_j) \prod_{1 \le j < k \le N} f_2(x_j,x_k).
\end{equation}
Three particular quantum many body systems of the Calogero-Sutherland
type are specified by the Schr\"odinger operators
\begin{subeqnarray}\label{schrops1}
H^{(H)} &=& - \sum_{j=1}^N {\partial^2 \over \partial x_j^2}
+{\beta^2 \over 4} \sum_{j=1}^N x_j^2 +\beta (\beta /2 - 1) \sum_{1 \le j < k \le N}
{1 \over (x_j - x_k)^2} \slabel{herm1}\\
H^{(L)} &=& -\sum_{j=1}^N {\partial^2
\over \partial x_j^2} + \sum_{j=1}^N\left( {\beta a' \over 2} \left( {a'\beta
\over 2} - 1 \right) {1 \over x_j^2} + {\beta^2 \over 4} x_k^2
\right) + 2 \beta(\beta / 2 - 1) \sum_{j,k=1 \atop j \ne k}^N {x_j^2 \over
(x_k^2 - x_j^2)^2} \slabel{lag1}\nonumber \\ \\
H^{(J)} &=&  - \sum_{j=1}^N {\partial^2
\over \partial \phi_j^2} + \sum_{j=1}^N\left( {a' \beta  \over 2} \left(
 {a'\beta
\over 2} - 1 \right) {1 \over \sin^2\phi_j} + {b'\beta \over 2} \left( {b'\beta
\over 2} - 1 \right) {1 \over \cos^2\phi_j} \right)\nonumber  \\
&&+ 2\beta(\beta/2 - 1) \sum_{j,k=1 \atop j \ne k}^N
 {\sin^2 \phi_j \cos^2 \phi_j 
\over (\sin^2\phi_j-\sin^2\phi_k) ^2}. \slabel{jacobi1}
\end{subeqnarray}
The superscripts $(H),\: (L), \: (J)$ stand for Hermite, Laguerre and Jacobi
respectively, and are chosen because of the relationship of these Schr\"odinger
operators to generalizations of the corresponding classical polynomials.

A direct calculation shows that there are eigenfunctions of the form
$e^{-\beta W/2}$ where
\begin{subeqnarray}\label{pots}
W^{(H)} &  = & {1 \over 2} \sum_{j=1}^N x_j^2 - \sum_{1 \le j < k \le N}
\log |x_k - x_j| \slabel{hermpot} \\
W^{(L)} & = & {1 \over 2} \sum_{j=1}^N x_j^2  - {a' \over 2}
 \sum_{j=1}^N \log x_j^2 - \sum_{1 \le j < k \le N}
\log |x_k^2 - x_j^2| \slabel{lagpot} \\
W^{(J)} & = &\! - {a' \over 2} \sum_{j=1}^N \log \sin^2 \phi_j -
{b'\over 2} \sum_{j=1}^N \log \cos^2 \phi_j -\!\!\!\sum_{1 \le j < k \le N} 
\!\!\log |\sin^2\phi_j - \sin^2\phi_k| \slabel{jacpot}
\end{subeqnarray}
Since these eigenfunctions are non-negative they correspond to the ground state
wavefunction $\psi_0$ (i.e.~they are the eigenfunctions with the most negative
eigenvalue $E_0$). Notice that $\psi_0$ is indeed of the type (\ref{gstate}).

Conjugation of the Schr\"odinger operators by 
the reciprocal of the ground state $e^{\beta W /2}$ gives the 
Fokker-Planck operators
\begin{equation}
{\cal L} := -{1 \over \beta} e^{-\beta W /2}(H- E_0) e^{\beta W /2}
= \sum_{j=1}^N {\partial \over \partial x_j} \Big (
{\partial W \over \partial x_j} + {1 \over \beta}
{\partial \over \partial x_j} \Big ) \label{fp-op}
\end{equation}
(for $H^{(J)}$ the coordinates $x_j$ are to replaced by $\phi_j$).  Thus the
Schr\"odinger equation
\begin{equation}
i {\partial \over \partial t} \psi (\{x_j\};t) = H  \psi (\{x_j\};t),
\label{schro}
\end{equation}
with $\psi = e^{-iE_0 t} e^{\beta W/2} P$ and $t = \tau / i \beta$ transforms
to the Fokker-Planck equation
\begin{equation}
{\partial \over \partial \tau} P = {\cal L} P.
\label{fp-eqn}
\end{equation}
The Fokker-Planck equation (\ref{fp-eqn}) describes the evolution of a 
classical gas in one-dimension with potential energy $W$ undergoing 
Brownian motion.

Two classes of problems associated with the Schr\"odinger operators 
(\ref{schrops1})
or equivalently the Fokker-Planck operator (\ref{fp-op}) with $W$ given by 
(\ref{pots}),
are the topic of this paper. The first is the discussion of some mathematical
properties relating to the eigenfunctions, while the second is the evaluation
of the density in the ground state and the exact solution of 
(\ref{fp-eqn}) for certain initial conditions.
These problems are in fact inter-related; we find that the
density for each system can be written in terms of a certain eigenstate and
that a summation theorem for the eigenstates gives an exact solution
of (\ref{fp-eqn}).

A feature of the  Schr\"odinger operators (\ref{schrops1}) is that after
conjugation with the ground state:
\begin{equation}
-e^{\beta W/2} (H-E_0) e^{-\beta W/2} = 
\sum_{j=1}^N \Big ( {\partial^2 \over \partial 
x_j^2} - \beta {\partial W \over \partial
x_j} {\partial \over \partial x_j} \Big ) \label{gaugeham}
\end{equation}
the resulting differential operator has a complete set of polynomial
eigenfunctions. In Section 2 we consider the form of the expansion
of these polynomials in terms of some different bases of symmetric
functions. We note that in the $N=1$ case, after a suitable change of 
variables, the operator (\ref{gaugeham}) with $W$ given by (\ref{pots}) is the
eigenoperator for the classical Hermite, Laguerre and Jacobi polynomials. 
Previous studies of the operator for general $N$ in the Jacobi case 
\cite{beer93a}
have established an orthogonality relation. Since the polynomials in
the Hermite and Laguerre cases are limiting cases of these generalized
Jacobi polynomials, we can obtain the corresponding orthogonality
relations via the limiting procedure.

The generalized Hermite polynomials, which are the polynomial eigenfunctions
of (\ref{fp-op}) with $W=W^{(H)}$ as given by (\ref{hermpot}), are 
studied in Section 3. Many
higher-dimensional analogues of properties of the classical Hermite
polynomials are obtained, including a generating function formula,
differentiation and integration formulas, a summation theorem and
recurrence relations.  An analogous study of the generalized
Laguerre polynomials is performed in Section 4. In Section 5 we relate
the problem of computing the ground state density for the
Schr\"odinger  operators (\ref{schrops1}) to the computation of particular
eigenstates. By using integral formulas for these eigenstates we are
able to compute the global density limit for even values of the
coupling $\beta$. In the case of the Schr\"odinger  operator (\ref{herm1}),
the limiting global density is the well known Wigner semi-circle
law. Also in Section 5, we give interpretation to results obtained in
Sections 3 and 4 for a summation formula. The interpretation is in 
terms of the solution of an initial value problem associated with the
Schr\"odinger equation (\ref{schro})

We conclude in Section 6 by identifying the formulas contained herein
which are to be found in previous works, and give reference to these
works (two of the most important references in this regard are
unpublished, handwritten manuscripts). In the Appendix we present
some results relating to generalized hypergeometric functions
depending on two sets of variables which are of relevance to the working
in Sections 3 and 4.

\setcounter{equation}{0}
\section{Inter-relationships}

Let us begin by explicitly calculating the operator (\ref{gaugeham}) for $W$
given by (\ref{pots}). In all cases it is
convenient to first change variables: for $W=W^{(H)}$ set $y_j :=
\sqrt{\beta / 2}\, x_j$, for $W=W^{(L)}$ set $y_j = \beta x_j^2 / 2$, while for
$W=W^{(J)}$ set
$y_j = \sin^2 \phi_j$. We then obtain
\begin{subeqnarray} \label{eops}
\tilde{H}^{( H)}&:=& -{2 \over  \beta}e^{\beta W^{(H)}/2} (H^{(H)} - E_0)
e^{-\beta W^{(H)}/2}  \nonumber\\
&=&\sum_{j=1}^N \Big (  {\partial^2 \over \partial y_j^2} -2 y_j
{\partial \over \partial y_j}+ {2\over \alpha} \sum_{k=1 \atop k\ne j}^N
{1 \over y_j - y_k} {\partial \over \partial y_j} \Big ),
\slabel{ehpoly} \\
\tilde{H}^{( L)} &:=&-{1 \over 2 \beta}e^{\beta W^{{(L)}}/2} 
(H^{{\rm (L)}} - E_0) e^{-\beta W^{{(L)}}/2} \nonumber\\
&=& \sum_{j=1}^N \Big ( y_j {\partial^2 \over \partial y_j^2} + (a - y_j +1)
{\partial \over \partial y_j}+ {2\over \alpha} \sum_{k=1 \atop k \ne j}^N
{y_j \over y_j - y_k} {\partial \over \partial y_j} \Big ),
\slabel{elpoly} \\
\tilde{H}^{(J)}&:=& -{1 \over 4} e^{\beta W^{(J)}/2}(H^{(J)} - E_0)  
e^{-\beta W^{(J)}/2} \nonumber \\  &=& \!\!
\sum_{j=1}^N\left( y_j (1 - y_j){\partial^2 \over \partial y_j^2} +
[{a} + 1 - y_j({a} + {b} + 2)]{\partial \over \partial
y_j}  +{2\over \alpha} \sum_{k=1 \atop k \ne j}^N
{y_j(1 - y_j) \over y_j - y_k}{\partial \over \partial y_j} \right)
\qquad\slabel{ejpoly} 
\end{subeqnarray}
where
$$
a := (\beta a' - 1)/2, \qquad b := (\beta b' - 1)/2, \qquad \alpha :=
2/\beta.
$$

In the one-variable case ($N=1$), these operators have 
a complete set of polynomial eigenfunctions given by the
classical  Hermite, Laguerre and Jacobi polynomials
\begin{subeqnarray}
H_n(y) &:=& n!\sum_{j=0}^{[n/2]} \frac{(-1)^j(2y)^{n-2j}}
{j!(n-2j)!} \slabel{her-1v} \\
L^a_{n}(y) &:=& {(a+1)_n \over n!} \sum_{j=0}^n \left ( {n \atop j} \right )
{(-y)^{j} \over (a+1)_j}
\slabel{lag-1v} \\
P_n^{(b,a)}(2y-1)&:=& (-1)^n\Big ( {n + a \atop n} \Big )
\sum_{j=0}^n {(n + a + b + 1)_j \over (a+1)_j} 
\Big ( {n \atop j} \Big ) (-y)^j
\end{subeqnarray}
respectively, where
$
(u)_n := u(u+1) \dots (u+n-1).
$

It is also true for general $N$ that there is a complete set of
polynomial eigenfunctions for each of the operators (\ref{eops}). This 
can be seen by computing their action on the monomial symmetric polynomial
$m_\kappa$, where $\kappa$ denotes a partition consisting of $N$ parts
$\kappa_j$. We obtain series of the form
\begin{subeqnarray}
e^{(H)}(\kappa,\alpha) m_\kappa & + &  \sum_{|\mu| <| \kappa|}
b^{(H)}_{\mu \kappa} m_\mu  \\
e^{(L)}(\kappa,\alpha) m_\kappa & + &  \sum_{|\mu| <| \kappa|}
b^{(L)}_{\mu \kappa} m_\mu \\
e^{(J)}(\kappa,\alpha) m_\kappa & + &  \sum_{\mu < \kappa}
a^{(J)}_{\mu \kappa} m_\mu\,  + \, \sum_{|\mu| <| \kappa|} 
b^{(J)}_{\mu \kappa} m_\mu
\end{subeqnarray}
respectively, where the notation $|\mu| < |\kappa|$ means 
$\sum_{j=1}^N \mu_j < \sum_{j=1}^N \kappa_j$, while the notation $\mu < \kappa$
means $\mu \ne \kappa$ but
$$
\sum_{j=1}^N \mu_j = \sum_{j=1}^N \kappa_j \quad {\rm and} \quad
\sum_{j=1}^p \mu_j \le \sum_{j=1}^p \kappa_j \quad {\rm for \: each}
\quad p=1,\dots, N
$$
Also, $a_{\mu \kappa}$, $b_{\mu \kappa}$ are coefficients independent of 
$y_j$ and 
\begin{equation}
e^{(H)}(\kappa, \alpha) = -2 |\kappa|, \qquad
e^{(L)}(\kappa, \alpha) = - |\kappa|
\end{equation}
(the explicit value of $e^{(J)}(\kappa, \alpha)$ can also be computed, however 
it is not needed in our subsequent discussion). This means there are
eigenfunctions of the form
\begin{subeqnarray}\label{sg.1}
\tilde{a}^{(H)}_{\kappa \kappa} m_\kappa & + &  \sum_{|\mu| <| \kappa|}
\tilde{b}^{(H)}_{\mu \kappa} m_\mu  \\
\tilde{a}^{(L)}_{\kappa \kappa} m_\kappa & + &  \sum_{|\mu| <| \kappa|}
\tilde{b}^{(L)}_{\mu \kappa} m_\mu \\
\tilde{a}^{(J)}_{\kappa \kappa} m_\kappa & + &  \sum_{\mu < \kappa}
\tilde{a}^{(J)}_{\mu \kappa} m_\mu \, +\,  \sum_{|\mu| <| \kappa|}
\tilde{b}^{(J)}_{\mu \kappa} m_\mu
\end{subeqnarray}
with eigenvalues $e^{(H)}(\kappa, \alpha)$, $e^{(L)}(\kappa, \alpha)$ and
$e^{(J)}(\kappa, \alpha)$ respectively.

Rather than study the eigenfunctions in the form (\ref{sg.1}), previous
studies \cite{macunp1,lass96a} have 
shown that it is advantageous to change basis
from the monomial symmetric polynomials to the Jack polynomials
\cite{stan89a,macunp1}. We recall that the Jack polynomial
$J_\kappa^{(\alpha)}(z_1,\dots, z_N)$ is the unique (up to normalization)
symmetric eigenfunction of the operator
\begin{equation}
D_2 := \sum_{j=1}^N z_j^2 {\partial^2 \over \partial z_j^2}
+ {2 \over \alpha} \sum_{j,k = 1 \atop j \ne k}^N
{z_j^2 \over z_j - z_k}{\partial \over \partial z_j}
\end{equation}
which has an expansion of the form
\begin{equation}
a_{\kappa \kappa} m_{\kappa} + \sum_{\mu<\ka} a_{\mu \kappa} m_{\mu}.
\label{jexpand}
\end{equation}
The notation $J_\kappa^{(\alpha)}$ is usually used for the particular
normalization $a_{(1^{|\kappa|})\kappa} = |\kappa|!$ in 
(\ref{jexpand}). However, for our purposes it is more convenient to choose a
different normalization, and to denote the corresponding Jack polynomial
by $C_\kappa^{(\alpha)}$ as in e.g.~\cite{kaneko93a}. This normalization is 
specified by requiring
\begin{equation} \label{mi.8}
(x_1 + \dots + x_N)^n =
\sum_{|\kappa| = n} C_\kappa^{(\alpha)}(x_1 , \dots , x_N)
\end{equation}
It is known (see e.g.~\cite{kaneko93a}) that $J_\kappa^{(\alpha)}$ and
$C_\kappa^{(\alpha)}$ are related by
\begin{equation} \label{farfel}
C_\kappa^{(\alpha)}(x_1 , \dots , x_N) =
 \alpha^{|\kappa|} |\kappa|! j_\kappa^{-1}
J_\kappa^{(\alpha)}(x_1 , \dots , x_N)
\end{equation}
where
\begin{subeqnarray}
j_{\ka} &:=& \prod_{s\in\ka} h_{\ka}^*(s)\: h^{\ka}_*(s)\slabel{mi.2}\\
\mbox{with}\hspace{2.7cm}
h_{\ka}^*(s) &:=& l_{\ka}(s) + \al(a_{\ka}(s)+1) \quad
h^{\ka}_*(s) := l_{\ka}(s) + 1 + \al a_{\ka}(s) \slabel{mi.3}
\hspace{1.5cm}
\end{subeqnarray}
In (\ref{mi.2}) and (\ref{mi.3}), $\kappa$ is regarded as 
a diagram, $s$ denotes a
node in the diagram and $a_\kappa(s)$ $(l_\kappa(s))$ denotes the arm
length (leg length) of the node (see e.g.~\cite{mac}).
In terms of the Jack polynomials, it is known \cite{macunp1,%
lass91a,lass91b,lass91c} that for each
partition $\kappa$ there is an eigenfunction of the form
\begin{subeqnarray}\label{mi.1}
H_\kappa(y_1,\dots,y_N;\alpha)& :=&
\sum_{\mu \subseteq \kappa} c_{\mu \kappa}^{(H)} 
C_\mu^{(\alpha)}(y_1,\dots,y_N)
\slabel{hexp} \\
L^a_\kappa(y_1,\dots,y_N;\alpha)& :=&
\sum_{\mu \subseteq \kappa} c_{\mu \kappa}^{(L)} 
C_\mu^{(\alpha)}(y_1,\dots
,y_N) \slabel{jexp}\\ \slabel{gexp}
G^{(a,b)}_\kappa(y_1,\dots,y_N;\alpha)& :=&
\sum_{\mu \subseteq \kappa} c_{\mu \kappa}^{(J)} 
C_\mu^{(\alpha)}(y_1,\dots,y_N) 
\end{subeqnarray}
where the notation $\mu \subseteq \kappa$ denotes 
$\mu_j \le \kappa_j$ for each $j=1,\dots,N$ and
$c_{\kappa \kappa} \ne 0$. These results can be
established by using known formulas for the action of the operators
\begin{subeqnarray}\label{defs.1}
E_k &:=& \sum_{i=1}^N x_i^k\,\frac{\partial}{\partial x_i} \\
D_k &:=& \sum_{i=1}^N x_i^k\,\frac{\partial^2}{\partial x_i^2} +
\frac{2}{\alpha}\sum_{i\neq j} \frac{x_i^k}{x_i-x_j}\frac{\partial}
{\partial x_i}
\end{subeqnarray}
for $k=0,1,2$ which for future reference we list here:
\begin{subeqnarray}\label{actions}
E_0 \frac{C^{(\alpha)}_{\kappa}(x)}{C^{(\alpha)}_{\kappa}(1^N)}
&=& \sum_{i=1}^N \bin{\kappa}{\kappa_{(i)}}\,
\frac{C^{(\alpha)}_{\kappa_{(i)}}(x)}{C^{(\alpha)}_{\kappa_{(i)}}
(1^N)} \slabel{id.1}\\
E_1 \ca_{\kappa}(x) &=& |\kappa| \ca_{\kappa}(x)\slabel{id.2} \\
E_2 \ca_{\kappa}(x) &=& \frac{1}{1+|\ka|}\sum_{i=1}^N
\bin{\kappa^{(i)}}{\kappa}\left(\kappa_i-\frac{i-1}{\alpha}
\right)\,\ca_{\kappa^{(i)}}(x) \slabel{id.3}\\
D_1 \frac{C^{(\alpha)}_{\kappa}(x)}{C^{(\alpha)}_{\kappa}(1^N)}
&=& \sum_{i=1}^N \bin{\kappa}{\kappa_{(i)}}\,\left(
\kappa_i-1+\frac{N-i}{\alpha}\right) \frac{C^{(\alpha)}_{\kappa_{(i)}}
(x)}{C^{(\alpha)}_{\kappa_{(i)}}(1^N)}  \slabel{id.4}\\
D_2 \ca_{\kappa}(x) &=& d_{\kappa} \ca_{\kappa}(x),
\quad d_{\kappa}:= \frac{2}{\alpha}|\kappa|(N-1) + \sum_{i=1}^N
\kappa_i\left(\kappa_i-1-\frac{2}{\alpha}(i-1)\right) \quad\slabel{id.5}
\end{subeqnarray}
(the action of $D_0$ can be computed from the commutator formula
$D_0 =  [E_0, D_1]$).
Here the generalized binomial coefficients $\left (
{\kappa \atop \sigma} \right )$ are defined by the expansion
\begin{equation}
{C_\kappa^{(\alpha)}(1+t_1, \dots, 1 + t_N) \over C_\kappa^{(\alpha)}(1^N)}
= \sum_{s=0}^{|\kappa|} \sum_{|\sigma| = s} \left (
{\kappa \atop \sigma} \right )
{C_\sigma^{(\alpha)}(t_1, \dots, t_N) \over C_\sigma^{(\alpha)}(1^N)}
\label{gbin}
\end{equation}
where $C_{\kappa}^{(\alpha)}(1^N)$ has the explicit form
\begin{equation}\label{ci.1}
C_\kappa^{(\alpha)}(1^N) = \frac{\al^{|\ka|}\,|\ka|!}
{j_{\ka}}\:\prod_{(i,j)\in\ka} \left( N-(i-1)+\al(j-1)\right).
\end{equation}
We have also used the notation
$$
\kappa_{(i)} := (\kappa_1, \dots, \kappa_{i-1}, \kappa_i - 1,
\kappa_{i + 1}, \dots, \kappa_N), \quad
\kappa^{(i)} := (\kappa_1, \dots, \kappa_{i-1}, \kappa_i + 1,
\kappa_{i + 1}, \dots, \kappa_N)
$$
(note that this is the opposite of what is used in \cite{muir82,kaneko93a}
but rather is that used by \cite{lass90a}).

The polynomials in (\ref{mi.1}) are referred to as 
generalized Hermite, Laguerre
and Jacobi polynomials respectively \cite{james75a}; they are 
uniquely specified up
to normalization as the eigenfunctions of the operators (\ref{eops}) with an
expansion in terms of Jack polynomials with highest weight (i.e.~largest
partition in reverse lexicographical ordering) $c_{\kappa \kappa}
C_\kappa^{(\alpha)}$.
For the normalization we choose
\begin{equation}
c_{\kappa \kappa}^{(H)} = 2^{|\kappa|} / C_\kappa^{(\alpha)}(1^N), \quad
c_{\kappa \kappa}^{(L)} = (-1)^{|\kappa|}/|\kappa|! C_\kappa^{(\alpha)}(1^N)
 \quad {\rm and} \quad
c_{\kappa \kappa}^{(J)} = 1. \label{coeff}
\end{equation}
With this choice, for $N=1$ the generalized Hermite and Laguerre polynomials 
exactly coincide with the classical Hermite and
Laguerre polynomials (\ref{her-1v}) and (\ref{lag-1v}) respectively, 
while in the $N=1$ case $G_{(k)}^{(a,b)}$ corresponds to the Jacobi 
polynomial $P_k^{(a,b)}(2y-1)$,  normalized so that the 
coefficient of $y^k$ is unity.

There have been a number of studies of the generalized Jacobi polynomials
$G_{\kappa}^{(a,b)}$ \cite{lass91c,debia87,macunp1}. 
In particular, it is known that these
polynomials are orthogonal with respect to the inner product
\begin{equation}
\langle f | g \rangle^{({\rm J})} := \prod_{l=1}^N \int_0^1 dy_l \,
y_l^a (1 - y_l)^b \prod_{1 \le j < k \le N} |y_k - y_j|^{2/\alpha}
f(y_1,\dots,y_N) g(y_1,\dots,y_N) \label{ijacobi}
\end{equation}
This is significant to the study of the generalized Hermite and Laguerre
polynomials, as both are limiting cases of the Jacobi polynomials. Thus by
comparing the operators $\tilde{H}^{(H)}$ and  $\tilde{H}^{(L)}$
with  $\tilde{H}^{(J)}$, and using the facts that the Jack polynomial 
$C_\kappa^{(\alpha)}$ is homogeneous of order $|\kappa|$ and that in the
expansion (\ref{gbin}) the binomial coefficient is non-zero if and only if
$\mu \subset \kappa$ \cite{kaneko93a}, we see that
\begin{equation}\label{hlim}
\lim_{b \to \infty}{2^{2 |\kappa|} (-b)^{|\kappa|} \over
C_\kappa^{(\alpha)}(1^N)}G_\kappa^{(b, b)}
\Big ({1 \over 2}(1 - {y_1 \over  b}), \dots, {1 \over 2}(1 - {y_N \over  b})
\Big ) = H_\kappa (y_1,\dots,y_N; \alpha)
\end{equation}
and
\begin{equation} \label{llim}
\lim_{b \to \infty}{ (-1)^{|\kappa|}b^{|\kappa|} \over
|\kappa|! C_\kappa^{(\alpha)}(1^N)} G_\kappa^{(a, b)}
(y_1/b, \dots, y_N/b; \alpha) =  L_\kappa^a(y_1,\dots,y_N; \alpha)
\end{equation}

It thus follows that by performing the same change of variables and
limiting procedure in  (\ref{ijacobi}), we will obtain inner products for which
the Hermite and Laguerre polynomials are orthogonal with respect to. We
find that for the generalized Hermite polynomials this inner product is
\begin{equation}
\langle f | g \rangle^{({\rm H})} := \prod_{l=1}^N \int_{-\infty}^\infty dy_l \,
  e^{-y_l^2} \prod_{1 \le j < k \le N} |y_k - y_j|^{2/\alpha}
f(y_1,\dots,y_N) g(y_1,\dots,y_N) \label{ih}
\end{equation}
while for the  generalized Laguerre polynomials it is
\begin{equation}
\langle f | g \rangle^{({\rm L})} := \prod_{l=1}^N \int_0^\infty dy_l \,
 y_l^a e^{-y_l} \prod_{1 \le j < k \le N} |y_k - y_j|^{2/\alpha}
f(y_1,\dots,y_N) g(y_1,\dots,y_N). \label{il}
\end{equation}
(these inner products have previously been identified by Lassalle
\cite{lass91a,lass91b}).

\setcounter{equation}{0}
\section{The generalized Hermite polynomials}
\subsection{The generating function}
The starting point and key source of inspiration in our studies of the
generalized Hermite and Laguerre polynomials is a private correspondence
with M.~Lassalle\cite{lass96a}, in which we 
received unpublished notes containing, amongst
other results, a  multi-variable generalization of the classical generating
function formula
\begin{equation}
\sum_{k=0}^\infty {H_k(y) z^k \over k!} = e^{2yz} e^{-z^2},
\end{equation}
which is given by the following result.

\vspace{.2cm}
\noindent
{\bf Proposition 3.1} \quad {\it Let $y := (y_1, \dots, y_N)$ and
$z := (z_1, \dots, z_N)$. The generalized Hermite polynomials $H_\kappa(y;
\alpha)$, defined in the previous section as polynomial eigenfunctions
of the operator (\ref{ehpoly}), which have highest weight term as in 
(\ref{hexp}) with the normalization specified 
by (\ref{coeff}), are given by the generating function
\begin{subeqnarray}
\sum_{\kappa} {1 \over |\kappa|!} H_\kappa(y;\alpha) C_\kappa^{(\alpha)}(z)
=  {}_0^{}{\cal F}_0^{(\alpha)}(2y;z) e^{-p_2(z)} \hspace{4cm}
\slabel{gfhermite} \\
\mbox{where} \hspace{14cm}\nonumber\\
{}_0^{}{\cal F}_0^{(\alpha)}(2y;z) :=
\sum_{\kappa}{1 \over |\kappa|!} { C_\kappa^{(\alpha)}(2y) C_\kappa^{(\alpha)}(z) \over  C_\kappa^{(\alpha)}(1^N)} \quad {\rm and} \quad
p_2(z) := \sum_{j=1}^N z_j^2. \qquad \slabel{f00}
\end{subeqnarray}
}

\vspace{.2cm}
A fundamental result in Lassalle's researches is an explicit formula for the
action of the operator $E_0^{(y)}$ (recall (\ref{id.1}); the superscript $(y)$
indicates operation with respect to the variables $y$) on
${}_0^{}{\cal F}_0^{(\alpha)}$:
\begin{equation} 
E_0^{(y)} \, {}_0^{}{\cal F}_0^{(\alpha)}(2y;z) = 
2 p_1(z)\:{}_0^{}{\cal F}_0^{(\alpha)}(2y;z),
\quad {\rm where} \quad p_1(z) := \sum_{j=1}^N z_j. \label{id.o}
\end{equation}
This formula follows from (\ref{id.1}) and the result 
\cite{kaneko93a,lass90a}
\begin{equation}
p_1(x)\,\ca_{\kappa}(x) = \frac{1}{1+|\ka|}\sum_{i=1}^N
\bin{\kappa^{(i)}}{\kappa} \ca_{\kappa^{(i)}}(x) \label{id.oo}
\end{equation}
Now in the notation of (\ref{defs.1}), the operator $\tilde{H}^{(H)}$ 
(\ref{ehpoly}) is given by
\begin{equation}
\tilde{H}^{(H)} = D_0 - 2 E_1 \label{hop}
\end{equation}
Knowledge of the action of $ D_0^{(y)}$ on ${}_0^{}{\cal F}_0^{(\alpha)}
(2y;z)$ is required to prove Proposition 3.1. Lassalle uses the
formulas (\ref{actions}) and (\ref{id.o}) to establish this action. 
We have observed that in
fact the required formula can be derived from (\ref{id.o}).  In our
derivation we make use of the
general fact that if $A^{(y)} F =
\hat{A}^{(z)} F$ and $B^{(y)} F =
\hat{B}^{(z)} F$,  then
\begin{equation}
A^{(y)}  B^{(y)} F =
A^{(y)}  \hat{B}^{(z)} F = \hat{B}^{(z)} A^{(y)} F =
 \hat{B}^{(z)} \hat{A}^{(z)} F
\label{commute}
\end{equation}
where the second equality follows because operators acting on different sets
of variables always commute.

\vspace{.2cm}
\noindent
{\bf Lemma 3.1} \quad {\it We have
\begin{subeqnarray}
D_1^{(y)}\:{}_0^{}{\cal F}_0^{(\alpha)}(2y;z) &=& 
\Big ( {2 \over \alpha}(N-1)p_1(z) + 2E_2^{(z)} \Big )
{}_0^{}{\cal F}_0^{(\alpha)}(2y;z) \slabel{mi.4} \\
D_0^{(y)}\:{}_0^{}{\cal F}_0^{(\alpha)}(2y;z) &=& 4 p_2(z) \, 
{}_0^{}F_0^{(\alpha)}(2y;z) \slabel{mi.5}
\end{subeqnarray} 
}

\vspace{.2cm}
\noindent
{\it Proof} \quad
Since $D_1^{(y)} = {1 \over 2} [E_0^{(y)}, D_2^{(y)}]$, using
(\ref{id.o}), the fact that $D_2^{(y)}$ is an eigenoperator for the Jack
polynomials, and (\ref{commute}) gives
\begin{eqnarray*}
D_1^{(y)}\:{}_0^{}{\cal F}_0^{(\alpha)}(2y;z
) & = & [ D_2^{(z)}, p_1(z)] \,
{}_0^{}{\cal F}_0^{(\alpha)}(2y;z
) \\
& = &
\Big ( {2 \over \alpha}(N-1)p_1(z) + 2E_2^{(z)} \Big )
{}_0^{}{\cal F}_0^{(\alpha)}(2y;z
)
\end{eqnarray*}
where the second equality follows by computing the commutator.
To derive the second result, note that  $D_0^{(y)} = [E_0^{(y)}, D_1^{(y)}]$,
so from (\ref{id.o}), (\ref{mi.4}) and (\ref{commute})
\begin{eqnarray*}
D_0^{(y)}\:{}_0^{}{\cal F}_0^{(\alpha)}(2y;z)
&=&[{2 \over \alpha}(N-1)p_1(z) + 2E_2^{(z)},2p_1(z)
] \,
{}_0^{}{\cal F}_0^{(\alpha)}(2y;z
) \nonumber \\
& = & 4p_2(z)\, {}_0^{}{\cal F}_0^{(\alpha)}(2y;z).
\end{eqnarray*}
\vspace{.2cm}

Let us now show how (\ref{mi.5}) is used in Lassalle's derivation of
(\ref{gfhermite}).

\vspace{.2cm}
\noindent
{\it Proof of Proposition 3.1} \quad We first want to show that $H_\kappa
(y;\alpha)$ as defined by the generating function (\ref{gfhermite}) is an
eigenfunction of (\ref{hop}) with eigenvalue $-2|\kappa|$. To do this, consider
the action of $E_1^{(z)}$ on both sides of (\ref{gfhermite}). On the
r.h.s.~we have
\begin{eqnarray}
\lefteqn{E_1^{(z)} \, {}_0^{}{\cal F}_0^{(\alpha)}(2y;z) e^{-p_2(z)}} 
\nonumber \\ &&
= e^{-p_2(z)} E_1^{(z)} \, {}_0^{}{\cal F}_0^{(\alpha)}(2y;z) - 2
p_2(z){}_0^{}{\cal F}_0^{(\alpha)}(2y;z) e^{-p_2(z)} \nonumber \\
&&= e^{-p_2(z)} E_1^{(y)} \, {}_0^{}{\cal F}_0^{(\alpha)}(2y;z) -
{1 \over 2} D_0^{(y)} \, {}_0^{}{\cal F}_0^{(\alpha)}(2y;z)  
e^{-p_2(z)} \nonumber
\\ &&
= - {1 \over 2} \sum_\kappa {1 \over |\kappa|!} (D_0^{(y)} - 2 E_1^{(y)})
H_\kappa(y;\alpha) C_\kappa^{(\alpha)}(z), \label{mi.6}
\end{eqnarray}
where the second equality follows by using (\ref{mi.5}) and noting
that since $E_1$ is an eigenoperator of the Jack polynomials, 
the definition (\ref{f00}) gives
$$
E_1^{(y)} \, {}_0^{}F_0^{(\alpha)}(2y;z) = E_1^{(z)} \, 
{}_0^{}F_0^{(\alpha)}(2y;z),
$$
and the final equality follows by substituting the generating function.
On the l.h.s., since $E_1^{(z)}$ is an eigenoperator of 
$C_\kappa^{(\alpha)}(z)$ with eigenvalue $|\kappa|$, from the 
definition (\ref{f00})
\begin{equation}
E_1^{(z)} \, \sum_{\kappa}{1 \over |\kappa|!} H_\kappa(y;\alpha)
 C_\kappa^{(\alpha)}(z) =  \sum_{\kappa}{1 \over |\kappa|!} H_\kappa(y;\alpha)
|\kappa | C_\kappa^{(\alpha)}(z) \label{mi.7}
\end{equation}
Equating coefficients of $ C_\kappa^{(\alpha)}(z)$ in (\ref{mi.6}) and
(\ref{mi.7}) shows that $H_\kappa(y;\alpha)$ is an eigenfunction of the
operator (\ref{hop}) with eigenvalue $-2|\kappa|$ as required.

It remains to check that $H_\kappa(y;\alpha)$ as given by (\ref{gfhermite})
has an expansion in terms of Jack polynomials with highest weight term
$2^{|\kappa|} C_\kappa^{(\alpha)}(y)/ C_\kappa^{(\alpha)}(1^N)$. This
follows from the fact that to compute the
coefficient of $ C_\kappa^{(\alpha)}(z)$ in ${}_0^{}{\cal F}_0^{(\alpha)}(2y;z) e^{-p_2(z)}$, the sum in (\ref{f00}) can be restricted to partitions with
modulus less than or equal to $|\kappa|$.

\vspace{.2cm}
The strategy of the above proof leads us to generalizations of the eigenvalue
equation. In the theory of Jack polynomials,
Macdonald \cite{mac} (see also \cite{seki77,debia87}) has given a family of
differential operators $\{D_N^j \}_{j=1,\dots,N}$ which have the Jack
polynomials as eigenfunctions, and for which the corresponding
eigenvalues are known explicitly. These operators are given by
\begin{eqnarray*}
D_N^p & := & \sum_{l=0}^{p} (\alpha)^{p-l} \sum_{1 \le i_1 < i_2 < \dots <
i_l} {1 \over \Delta_+} \Big (z_{i_1} {\partial \over \partial z_{i_1}} \dots
z_{i_l}{\partial \over \partial z_{i_l}}\Big ) \Delta_+ \\
& & \times \sum_{1 \le i_{l+1} < \dots < i_p \le N \atop \ne i_1 \dots \ne i_l}
\Big ( z_{i_{l+1}}{\partial \over \partial z_{i_{l+1}}} \dots z_{i_p}
{\partial \over \partial z_{i_p}} \Big ),
\end{eqnarray*}
where  $\Delta_+ := \prod_{1 \le j < k \le N} (z_k - z_j)$.
Furthermore, if we define the  corresponding generating function by
$$
 D_N(X; \alpha ) := \sum_{k=0}^N X^{N-k} D_N^k
$$
then the eigenvalues are given by
\begin{equation} 
e(\kappa,\alpha;X) := \prod_{j=1}^N (X + N - j + \alpha \kappa_j).
\label{gev} 
\end{equation}

The operator $E_1$ is related to the Macdonald operator $D_N^1$ by
$D_N^1 = \alpha E_1 + N(N-1)/2$. By considering the analogue of
(\ref{mi.6}) with $E_1^{(z)}$ replaced by $D_N^j$
$(j=1,2,\dots,N)$, a family of differential operators which have
the generalized Hermite polynomials as eigenfunctions can be given
The r.h.s.~of (\ref{mi.6}) is then computed according to the
Baker-Campbell-Hausdorff formula
\begin{eqnarray}\lefteqn{
D_N^{j\,(z)}\,f \,e^{p_1(z)} = e^{p_1(z)}\left(D_N^{j\,(z)}
+\left[D_N^{p(z)},p_1(z)\right]  \right.} \nonumber \\&&
\left.+\frac{1}{2!}\left[\left[D_N^{j\,(z)},p_1(z)\right],p_1(z)\right] +
\cdots \frac{1}{n!}\left[\cdots\left[D_N^{j\,(z)},p_1(z)\right],\cdots,
p_1(z)\right] \right) f
\end{eqnarray}
Note that the sum on the r.h.s.~terminates after the $n$-th nested
commutator since the highest derivative in $D_N^{j(z)}$ has degree
$j$.

Following the derivation of the eigenvalue equation given in
in the proof of Proposition 3.1, and thus using (\ref{id.o}), (\ref{commute})
and the fact that since $D_N^{j}$ is an eigenoperator of the Jacks we have
$$
D_N^{j\,(z)} {}_0^{}{\cal F}_0^{(\alpha)}(2y;z) = 
D_N^{j\,(y)} {}_0^{}{\cal F}_0^{(\alpha)}(2y;z)
$$
we can immediately deduce a family of $N$ independent eigenoperators
of the polynomials $H_\kappa(y;
\alpha)$, together with the corresponding eigenvalues.

\vspace{.2cm}
\noindent
{\bf Proposition 3.2} \quad {\it Let
\begin{eqnarray*}\
\tilde{H}_j^{(H)} &:=& D_N^{j(y)}    - 
\frac{1}{4}\left[D_0^{(y)}, D_N^{j(y)} \right] + \frac{1}{4^2\,2!}
\left[D_0^{(y)},\left[D_0^{(y)},  D_N^{j(y)} \right]\right] - \cdots
\\
&&+\:\frac{(-1)^n}{4^n\,n!}\left[D_0^{(y)},\left[D_0^{(y)},\cdots,\left[
D_0^{(y)}, D_N^{j(y)} \right] \right]\cdots\right]
\end{eqnarray*}
We have that $H_\kappa(y;\alpha)$ is an eigenfunction of $\tilde{H}_j^{(H)}$
for each $j=1,\dots,N$, with eigenvalue $e_j(\kappa;\alpha)$ given by
the coefficient of $X^{N-j}$ in (\ref{gev}).
}

\subsection{Consequences of the generating function}
The generating function formula (\ref{gfhermite}) can be used to deduce
higher-dimensional analogues of the classical properties of the Hermite
polynomials
\begin{subeqnarray}
H_n(-y) &=& (-1)^n\,H_n(y) \slabel{h.1}\\
\frac{d}{dy}\,H_n(y) &=& 2nH_{n-1}(y) \slabel{h.3}\\
2yH_n(y) &=& H_{n+1}(y) + 2nH_{n-1}(y). \slabel{h.4}
\end{subeqnarray}

\vspace{.2cm}
\noindent
{\bf Proposition 3.3} \quad {\it We have
$$
H_{\ka}(-y;\alpha) = (-1)^{|\ka|}\,H_{\ka}(y;\alpha).
$$
}

\vspace{.2cm}
\noindent
{\it Proof} \quad Replacing $y$ by $-y$ in (\ref{gfhermite}) and using the
fact that $ {}_0^{}{\cal F}_0^{(\alpha)}(2\mu y;z) =
 {}_0^{}{\cal F}_0^{(\alpha)}(2y;\mu z)$, where $\mu$ is any scalar, and
that $ C_\kappa^{(\alpha)}$ is homogeneous of order $|\kappa|$, gives
$$
\sum_{\kappa} {1 \over |\kappa|!} H_\kappa(-y;\alpha) C_\kappa^{(\alpha)}(z)
=  {}_0^{}{\cal F}_0^{(\alpha)}(-2y;z) e^{-p_2(z)}=
{}_0^{}{\cal F}_0^{(\alpha)}(2y;-z) e^{-p_2(-z)}
$$
$$
= \sum_{\kappa} {1 \over |\kappa|!}(-1)^{|\kappa|}H_\kappa(y;\alpha) C_\kappa^{(\alpha)}(z).
$$
The result follows by equating coefficients of $C_\kappa^{(\alpha)}(z)$.
\vspace{.2cm}

\noindent
{\bf Proposition 3.4} \quad {\it We have
$$
E_0^{(y)}\, H_{\ka}(y;\alpha) = 2\sum_{i=1}^N \bin{\ka}{\ka_{(i)}}\,
H_{\ka_{(i)}}(y;\alpha)
$$
}

\vspace{.2cm}
\noindent {\it Proof} \quad Applying $E_0^{(y)}$ to the generating function
(\ref{gfhermite}) and using (\ref{id.o}) gives
\begin{eqnarray}
\sum_{\kappa} {1 \over |\kappa|!} 
E_0^{(y)} \, H_\kappa(y;\alpha)  C_\kappa^{(
\alpha)}(z) & = & 2 p_1(z) {}_0^{}{\cal F}_0^{(\alpha)}(2y;z) e^{-p_2(z)}
\nonumber \\
& = &  2 p_1(z) \sum_{\kappa} {1 \over |\kappa|!} 
\, H_\kappa(y;\alpha) C_\kappa^{(
\alpha)}(z) \label{abov}
\end{eqnarray}
Using the formula (\ref{id.oo}) the final formula on the r.h.s.~of (\ref{abov})
can be rewritten as
\begin{equation}
2 \sum_{\kappa} {1 \over (1 + |\kappa|)!} \sum_{i=1}^N \Big ({\kappa^{(i)} \atop
\kappa} \Big )  H_{\ka}(y;\alpha) C_{\kappa^{(i)}}^{(
\alpha)}(z)= 2 \sum_{\kappa} {1 \over |\kappa|!}  \sum_{i=1}^N
\bin{\ka}{\ka_{(i)}}\,
H_{\ka_{(i)}}(y;\alpha) C_\kappa^{(
\alpha)}(z) \label{abov1}
\end{equation} 
Equating coefficients of $C_\kappa^{(
\alpha)}(z)$ on the l.h.s.~of (\ref{abov}) and on the r.h.s.~of
(\ref{abov1}) gives the stated result.

\vspace{.2cm}
\noindent
{\bf Proposition 3.5} \quad {\it We have 
$$
2p_1(y) H_{\ka}(y;\alpha) = \al\sum_i^N\bin{\ka^{(i)}}{\ka}\frac{j_{\ka}}
{j_{\ka^{(i)}}} \left( N-i+1 +\al\ka_i\right) H_{\ka^{(i)}}(y)
+ 2\sum_i\bin{\ka}{\ka_{(i)}} H_{\ka_{(i)}}(y)
$$
}

\vspace{.2cm}
\noindent
{\it Proof} \quad From the generating function (\ref{gfhermite}) we have
\begin{eqnarray}
2 \sum_{\kappa} {1 \over |\kappa|!} p_1(y) H_\kappa(y;\alpha) C_\kappa^{(
\alpha)}(z) &  = & 2 p_1(y)  {}_0^{}{\cal F}_0^{(\alpha)}(2y;z) e^{-p_2(z)}
 =  \Big ( E_0^{(z)} \,  {}_0^{}{\cal F}_0^{(\alpha)}(2y;z) \Big )
 e^{-p_2(z)} 
\nonumber \\ & & =
  \Big ( E_0^{(z)} \,  {}_0^{}{\cal F}_0^{(\alpha)}(2y;z) 
 e^{-p_2(z)} \Big ) + 2p_1(z)  {}_0^{}{\cal F}_0^{(\alpha)}(2y;z) e^{-p_2(z)}
\nonumber \\ &&
= \Big ( E_0^{(z)} + 2p_1(z) \Big ) \sum_{\kappa} {1 \over |\kappa|!}
 H_{\ka}(y;\alpha) C_\kappa^{(
\alpha)}(z) \label{abov2}
\end{eqnarray}
Using the formula (\ref{id.1}) and writing 
$$
2 p_1(z) \sum_{\kappa} {1 \over |\kappa|!}
 H_{\ka}(y;\alpha) C_\kappa^{(
\alpha)}(z)
$$
as in the proof of Proposition 3.4 shows that we can rewrite the last
expression on the r.h.s.~of (\ref{abov2}) as
$$
\sum_{\kappa} {1 \over |\kappa|!}  C_\kappa^{\alpha}(1^N)
\sum_{i=1}^N \bin{\ka}{\ka_{(i)}}H_{\ka}(y;\al)\,{C_{\kappa_{(i)}}^{(
\alpha)}(z) \over C_{\kappa_{(i)}}^{(
\alpha)}(1^N)} + 2  \sum_{\kappa} {1 \over |\kappa|!}
\sum_{i=1}^N\bin{\ka}{\ka_{(i)}}\,
H_{\ka_{(i)}}(y;\alpha) C_\kappa^{(
\alpha)}(z)
$$
$$
=  \sum_{\kappa} {1 \over (1+|\kappa|)!}\sum_{i=1}^N \bin{\ka}{\ka^{(i)}}
H_{\ka^{(i)}}(y;\al)\,{C_{\kappa^{(i)}}^{(
\alpha)}(1^N)\over C_\kappa^{(\alpha)}(1^N)}  C_\kappa^{(
\alpha)}(z)
+  2  \sum_{\kappa} {1 \over |\kappa|!}
\sum_{i=1}^N\bin{\ka}{\ka_{(i)}}\,
H_{\ka_{(i)}}(y;\alpha) C_\kappa^{(
\alpha)}(z)
$$
The stated formula now follows by equating coefficients of $ C_\kappa^{(
\alpha)}(z)$ on the l.h.s.~of (\ref{abov2}) and the r.h.s.~of the above
equation, and using (\ref{ci.1})) to rewrite 
${C_{\kappa^{(i)}}^{(\alpha)}(1^N)/C_\kappa^{(\alpha)}(1^N)}$.

Another consequence of the generating function  (\ref{gfhermite}) relates
to an analogue of the formula (\ref{mi.8}).

\vspace{.2cm}
\noindent
{\bf Proposition 3.6} \quad {\it We have
$$
H_k({1 \over \sqrt{N}} p_1(y) ) = N^{-k/2}\,\sum_{|\kappa| = k}
H_\kappa(y;\alpha) C_\kappa^{(\alpha)}(1^N).
$$
}

\vspace{.2cm}
\noindent
{\it Proof} \quad Set $z_1 = \cdots =z_N = c$ in (\ref{gfhermite}) and
note that
\begin{equation} \label{dia.1}
{}_0^{}{\cal F}_0^{(\alpha)}(2y;c,\ldots,c) = \sum_\kappa
{c^{|\kappa|} \over |\kappa|!} C_\kappa^{(\alpha)}(2y) =
e^{2cp_1(y)}
\end{equation}
(the last equality follows from (\ref{mi.8})) to conclude
$$
\sum_{\kappa} {c^{|\kappa|} \over |\kappa|!} H_\kappa(y;\alpha)
 C_\kappa^{(\alpha)}(1^N) = e^{2cp_1(y)} e^{-Nc^2} = \sum_{k=0}^\infty
{c^k \over N^{k/2}\,k!}\, H_k({1 \over \sqrt{N}} p_1(y) ).
$$
The result now follows by equating coefficients of $c^k$.

\vspace{.2cm}
Notice that each term on the r.h.s.~of 
the above formula is an eigenfunction of the operator (3.5)
with eigenvalue $-2|\kappa|$. Thus $ H_k({1 \over \sqrt{N}} p_1(y) )$ is 
also an eigenfunction of (\ref{hop}) with eigenvalue $-2k$. This latter 
fact can be checked
directly, and has been observed previously \cite{gura96}.

\subsection{Integration formulas}
Using the generating function (\ref{gfhermite}) and the orthogonality of the
generalized Hermite polynomials with respect to the inner product
(\ref{ih}), a number of integration formulas can be obtained. In
particular, we can obtain the multidimensional analogues of the classical
formulas
\begin{subeqnarray}\label{hint.0}
\int_{-\infty}^\infty dy \, e^{-y^2} \Big ( H_k (y) \Big )^2 & = &
\sqrt{\pi} 2^{k} k! \slabel{hint.1}\\
{2^{-k} \over \sqrt{\pi}} \int_{-\infty}^\infty dy \, e^{-y^2}
H_k(y+x)& =&  x^k \slabel{hint.2}\\
{2^k \over \sqrt{\pi}} \int_{-\infty}^\infty dy \, e^{-y^2}
(x + iy)^k & = & H_k(x).  \slabel{hint.3}
\end{subeqnarray}
To present these analogues let us introduce the notation
\begin{equation}
d\mu^{(H)}(y) := \prod_{j=1}^N e^{-y_j^2} \prod_{1 \le j < k \le N}
|y_j - y_k|^{2/\alpha} \, dy_1 \dots dy_N.\label{hm}
\end{equation}

\vspace{.2cm}
\noindent
{\bf Proposition 3.7} \quad  {\it We have
$$
{\cal N}_\kappa^{(H)} := \int_{(-\infty, \infty)^N}
\Big (  H_\kappa(y;\alpha) \Big )^2 d\mu^{(H)}(y) = {2^{|\kappa|}
|\kappa|! {\cal N}_0^{(H)} \over  C_\kappa^{(\alpha)}(1^N)}
$$
where
$$
{\cal N}_0^{(H)} :=  \int_{(-\infty, \infty)^N} d\mu^{(H)}(y) =
2^{-N(N-1)/2\alpha}
\pi^{N/2} \prod_{j=0}^{N-1}{\Gamma(1+(j+1)/\alpha) \over
 \Gamma (1 + 1/\alpha )}.
$$
}

\vspace{.2cm}
\noindent
{\it Proof} \quad Multiplying both sides of the generating function
(\ref{gfhermite}) by $H_\kappa(y;
\alpha)$, integrating with respect to the measure (\ref{hm}), and using
the orthogonality property of $\{H_\kappa(y;\alpha) \}_\kappa$ with
respect to the inner product (\ref{ih}) gives
\begin{equation}
{ {\cal N}_\kappa^{(H)} \over |\kappa|!}  C_\kappa^{(\alpha)}(z) =
e^{-p_2(z)}  \int_{(-\infty, \infty)^N}{}_0^{} {\cal F}_0^{(\alpha)}(2y;z)
H_\kappa(y;\alpha) \,  d\mu^{(H)}(y).
\label{abov3}
\end{equation}
Set $z_1 = \dots = z_N = c$, substitute (\ref{dia.1})  
in the r.h.s.~of (\ref{abov3}) and complete the
square to show
$$
{ {\cal N}_\kappa^{(H)} \over |\kappa|!}  C_\kappa^{(\alpha)}(c) 
=  \int_{(-\infty, \infty)^N} e^{-p_2(y)} \prod_{1 \le j < k \le N}
|y_k - y_j|^{2/\alpha} H_\kappa(y+c;\alpha) \, dy_1 \dots dy_N.
$$
Now take the limit $c \to \infty$. Since from (\ref{gbin}) and (\ref{hexp})
$$
\lim_{c \to \infty} { H_\kappa(y+c;\alpha) \over C_\kappa^{(\alpha)}(c)} =
{2^{|\kappa|} \over C_\kappa^{(\alpha)}(1^N)}
$$
the stated formula for $ {\cal N}_\kappa^{(H)} $ follows. The formula for 
$ {\cal N}_0^{(H)}$ is a well known limiting case of Selberg's integral.

The analogues of (\ref{hint.2}) and (\ref{hint.3}) can be derived from
the following integration formula.
\vspace{2mm}\\
{\bf Proposition 3.8}\quad {\it We have
$$
\int_{(\infty,\infty)^N} {}_0{\cal F}_0^{(\alpha)}(2y;z)
{}_0{\cal F}_0^{(\alpha)}(2y;w)\: d\mu(y) =
e^{p_2(w)+p_2(z)}\,{\cal N}_0^{(H)}\,{}_0{\cal F}_0^{(\alpha)}(2z;w)
$$
}

\vspace{2mm}
\noindent
{\it Proof}\quad Substitute the generating function (\ref{gfhermite}) for
${}_0{\cal F}_0^{(\alpha)}$ and integrate term-by-term using the
orthogonality property of the $\{H_{\kappa}(y;\alpha)\}_{\kappa}$ 
with respect to the inner product (\ref{ih})
and the normalization integral of Proposition 3.7. The resulting
series is identified as ${}_0{\cal F}_0^{(\alpha)}(2z;w)$ according
to the definition (\ref{f00}).
\vspace{2mm}\\
{\bf Corollary  3.1}\quad {\it We have
$$
\int_{(-\infty,\infty)^N} {}_0^{}{\cal F}_0^{(\alpha)}(2y;z)
H_{\kappa}(y,\alpha) \: d\mu(y) = e^{p_2(z)}\,{\cal N}_0^{(H)}
\frac{2^{|\kappa|}C^{(\alpha)}_{\kappa}(z)}{C^{(\alpha)}_{\kappa}(1^N)}
$$
}
{\it Proof}\quad Multiply both sides of the integration formula of
Proposition 3.8 by $\exp(-p_2(w))$ and substitute for
$\exp(-p_2(w)){}_0{\cal F}^{(\alpha)}_0(2y;w)$ using (\ref{gfhermite}). 
The result follows by equating coefficients of $C^{(\alpha)}_{\kappa}(w)$
on both sides.
\vspace{2mm}\\
{\bf Corollary  3.2}\quad {\it We have
$$
e^{-p_2(z)}\,H_{\kappa}(z;\alpha) = \frac{2^{|\kappa|}}{{\cal N}_0^{(H)}
C^{(\alpha)}_{\kappa}(1^N)} \int_{(\infty,\infty)^N} 
{}_0{\cal F}_0^{(\alpha)}(2y;-iz) C^{(\alpha)}_{\kappa}(iy)
\:d\mu(y)
$$
}
{\it Proof}\quad By writing $iw$ for $w$ in Proposition 3.8, we can
replace ${}_0{\cal F}^{(\alpha)}_0(2y;w)$ by \linebreak
${}_0{\cal F}^{(\alpha)}_0(2iy;w)$, ${}_0{\cal F}^{(\alpha)}_0(2z;w)$ 
by ${}_0{\cal F}^{(\alpha)}_0(2iz;w)$ and $\exp(-p_2(w))$ 
by $\exp(p_2(w))$. The result follows by using the generating function
(\ref{gfhermite}) to substitute for 
$\exp(-p_2(w)){}_0{\cal F}^{(\alpha)}_0(2iz;w)$
and equating coefficients of $C^{(\alpha)}_{\kappa}(w)$.

The integration formula of Corollary 3.2 can be used to derive the
analogue of the classical summation formula \cite{erdelyi2}
$$
\sum_{k=0}^{\infty}\frac{H_k(w)H_k(z)}{k! 2^k \sqrt{\pi}} \;t^k
=\frac{1}{\sqrt{\pi}} (1-t^2)^{-1/2} e^{-t^2(z^2+w^2)/(1-t^2)}
e^{2wzt/(1-t^2)},\quad |t|<1.
$$
\vspace{2mm}\\
{\bf Proposition 3.9}\quad {\it For $|t|<1$ we have
\begin{eqnarray*}
G^{(H)}(w,z;t)
 &:=& \sum_{\kappa} \frac{H_{\kappa}(w;\alpha)H_{\kappa}(z;\alpha)}
{ {\cal N}_{\kappa}^{(H)}} t^{|\kappa|} \\
&=& \frac{1}{{\cal N}_0^{(H)}} (1-t^2)^{-Nq/2}\;
\exp\left(-\frac{t^2}{(1-t^2)}\left(p_2(z) + p_2(w)\right)\right)
\\ && \times
{}_0^{}{\cal F}^{(\alpha)}_0\left(
\frac{2wt}{(1-t^2)^{1/2}};\frac{z}{(1-t^2)^{1/2}}\right )
\end{eqnarray*}
where $q=1+(N-1)/\alpha$. }\\
{\it Proof}\  Substituting the integral representation of Corollary
3.2 for $H_{\kappa}(z;\alpha)$ and $H_{\kappa}(w;\alpha)$ in the 
definition $G^{(H)}(w,z;t)$, we see that the sum over $\kappa$ can be 
recognized in terms of ${}_0{\cal F}^{(\alpha)}_0$ and thus
\begin{eqnarray*}
\lefteqn{
G^{(H)}(w,z;t) = e^{p_2(z)+p_2(w)}\frac{1}{({\cal N}_0^{(H)})^3} } \\
&& \times
\int_{(-\infty,\infty)^N} d\mu^{(H)}(y_a)\int_{(-\infty,\infty)^N} 
d\mu^{(H)}(y_b)
{}_0{\cal F}^{(\alpha)}_0(2y_a;-iw){}_0{\cal F}^{(\alpha)}_0(2y_b;-iz)
{}_0{\cal F}^{(\alpha)}_0(2y_a;-ty_b)
\end{eqnarray*}
We now use Proposition 3.8 to integrate over $y_a$. This gives
\begin{eqnarray*}
G^{(H)}(w,z;t)& =& e^{p_2(z)}\frac{1}{({\cal N}_0^{(H)})^2} 
\int_{(-\infty,\infty)^N} d\mu^{(H)}(y_b) {}_0{\cal F}^{(\alpha)}_0(2y_b;-iz)
{}_0{\cal F}^{(\alpha)}_0(2iw;ty_b)\,e^{t^2p_2(y_b)}\\
& =& e^{p_2(z)}\frac{1}{({\cal N}_0^{(H)})^2} 
(1-t^2)^{-(N/2 +N(N-1)/2\alpha)}\\
&\times &
\int_{(-\infty,\infty)^N} d\mu^{(H)}(y_b) {}_0{\cal F}^{(\alpha)}_0(2y_b
;-iz(1-t^2)^{-1/2}) {}_0{\cal F}^{(\alpha)}_0(2y_b
;iwt(1-t^2)^{-1/2})
\end{eqnarray*}
where the second equality follows by combining $d\mu^{(H)}(y_b)$ and
$\exp(t^2p_2(y_b))$ (recall (\ref{hm})) and changing variables.
The integration over $d\mu^{(H)}(y_b)$ can now be performed using
Proposition 3.8, and the summation formula for $G^{(H)}(w,z;t)$ results.

\vspace{2mm}
Notice from (\ref{dia.1}) that in the special case that $w_1=\cdots
=w_N=c$, the summation formula of Proposition 3.9 is entirely in
terms of elementary functions:
\begin{equation} \label{cat.1}
G^{(H)}(w,z;t) = \frac{1}{{\cal N}_0^{(H)}}
 (1-t^2)^{-Nq/2}
\exp\left(-\frac{1}{(1-t^2)}\sum_{j=1}^N(t^2 z_j^2 
- 2tcz_j + t^2 c^2) \right).
\end{equation}
Interpretation of this result in terms of an
explicit solution of the Fokker-Planck equation (\ref{fp-eqn}) 
with $W$ given by (\ref{hermpot}) will be discussed in Section 5.

In Corollary 3.2 a certain integral transform is applied to the Jack
polynomial to obtain the generalized Hermite polynomial. It has been
observed by Lassalle that the generalized Hermite polynomials can be
obtained from the Jack polynomials by the action of a certain
exponential differential operator. Thus from the formula (\ref{mi.5}), 
we see that
$$
\Big ( {1 \over 4} D_0^{(y)} \Big )^k {}_0{\cal F}^{(\alpha)}_0
(2y;z)= \Big ( p_2(z) \Big )^k {}_0^{}{\cal F}^{(\alpha)}_0
(2y;z),
$$
which after multiplication by $(-1)^k/k!$ and summing over $k$ gives
$$
\exp \Big (- {1 \over 4} D_0^{(y)} \Big )
{}_0{\cal F}^{(\alpha)}_0
(2y;z) = e^{-p_2(z)} {}_0^{}{\cal F}^{(\alpha)}_0
(2y;z).
$$
Use of the generating function (\ref{gfhermite}) on the r.h.s.~and equating
coefficients of $C_\kappa^{(\alpha)}(z)$ gives Lassalle's formula
\begin{equation} \label{china.1}
{2^{|\kappa|} \over C_\kappa^{(\alpha)}(1^N)} 
\exp \Big (- {1 \over 4} D_0^{(y)} \Big )C_\kappa^{(\alpha)}(y)
= H_\kappa(y;\alpha).
\end{equation}
Comparison with the formula of
Corollary 3.2,  and use of the fact that $\{ C_\kappa^{(\alpha)}(y)
\}_\kappa$ forms a basis for  symmetric analytic functions shows that
for any symmetric analytic function $f(y)$,
\begin{equation} \label{china.2}
{e^{p_2(z)} \over {\cal N}_0^{(H)}} \int_{(-\infty, \infty)^N}
{}_0^{}{\cal F}^{(\alpha)}_0
(2y;-iz) f(iy) \, d \mu^{(H)}(y) =
\exp \Big ( -{1 \over 4} D_0^{(z)} \Big ) f(z).\label{ci.2}
\end{equation}
{}From (\ref{ci.2}) we see that if
\begin{subeqnarray}
F(z) &=& {e^{p_2(z)} \over {\cal N}_0^{(H)}} \int_{(-\infty, \infty)^N}
{}_0^{}{\cal F}^{(\alpha)}_0
(2y;-iz) f(iy) \, d \mu^{(H)}(y) \hspace{20mm}\slabel{ci.5}\\
\!\!\!\mbox{then}\hspace{4cm} && \nonumber\\
f(z) &=& \exp \Big ( {1 \over 4} D_0^{(z)} \Big ) F(z).\slabel{ci.3}
\end{subeqnarray}
On the other hand, by replacing $z$ by $iz$ and $f(ix)$ by
$F(x)$ we have
\begin{equation}
{e^{-p_2(z)} \over {\cal N}_0^{(H)}} \int_{(-\infty, \infty)^N}
{}_0^{}{\cal F}^{(\alpha)}_0 (2y;z) F(y) \, d \mu^{(H)}(y) =  
\exp \Big ( {1 \over 4} D_0^{(z)} \Big ) F(z).\label{ci.4}
\end{equation}
Comparison of (\ref{ci.3}) and (\ref{ci.4}) gives
\begin{equation} \label{china.7}
f(z) = {e^{-p_2(z)} \over {\cal N}_0^{(H)}} \int_{(-\infty, \infty)^N}
{}_0^{}{\cal F}^{(\alpha)}_0
(2y;z) F(y) \, d \mu^{(H)}(y)
\end{equation}
which is the inversion formula for the transform (\ref{ci.5}) (in the case
$\alpha = \infty$ (\ref{ci.5}) corresponds to the Fourier transform).

\setcounter{equation}{0}
\section{The generalized Laguerre polynomials}

The generalized Laguerre polynomials, defined as the polynomial
eigenfunctions of the operator (\ref{elpoly}) of the form (\ref{jexp}) with
normalization (\ref{coeff}), also satisfy higher-dimensional analogues of
their classical counterparts. A number of these formulas have been proved in
the case $\alpha = 2$ by Muirhead \cite{muir82} and for general $\alpha$ but 
$N=2$ by Yan \cite{yan92a}. Below we will develop the theory of generalized 
Laguerre
polynomials by presenting the analogues of the classical generating functions,
the series expansion (\ref{lag-1v}), recurrence and differentiation formulas,
integration formulas and a summation formula. In Section 6 we will
identify the formulas known to Muirhead and Yan, as well as those which
can be found in the work of Lassalle and Macdonald.

\subsection{Generating functions}
The classical Laguerre polynomials can be defined by either of the
generating functions
\begin{subeqnarray}
e^z {J_{a}(2\sqrt{yz}) \over (yz)^{a/2}} &=& \sum_{n=0}^\infty
{1 \over \Gamma (n + a + 1)} L_n^a(y) z^n
\hspace{2cm}\slabel{l-gen1} \\
\mbox{or}\hspace{8cm} &&\nonumber\\
(1-z)^{-(a+1)} e^{yz/(z-1)} &=& \sum_{n=0}^\infty L_n^a(y) z^n 
\hspace{2cm}\slabel{l-gen2}
\end{subeqnarray}
where in (\ref{l-gen1}) $J_a$ denotes the Bessel function.
These generating functions have the following higher-dimensional analogues.

\vspace{.2cm}
\noindent
{\bf Proposition 4.1} \quad {\it We have
\begin{equation}
e^{p_1(z)} {}_0^{}{\cal F}_1^{(\alpha)}(a + q;
x;-z) = \sum_{\kappa} {L_\kappa^a(x;\alpha)  C_\kappa^{(\alpha)}(z)
\over [a + q]_\kappa^{(\alpha)}}
\label{l-gen3}
\end{equation}
where
\begin{subeqnarray}
q := 1 + (N-1)/\alpha, \hspace{6cm}\slabel{qq} \\ \slabel{recordar}
{}_p^{}{\cal F}_r^{(\alpha)}(a_1,\dots, a_p;b_1,\dots,b_r;x;z) := 
\sum_{\kappa} {1 \over |\kappa|!}
{[a_1]_\kappa^{(\alpha)} \dots [a_p]_\kappa^{(\alpha)} \over
[b_1]_\kappa^{(\alpha)} \dots [b_r]_\kappa^{(\alpha)} }
{ C_\kappa^{(\alpha)}(x) C_\kappa^{(\alpha)}(z) \over
 C_\kappa^{(\alpha)}(1^N) }
\end{subeqnarray}
with $ 
[c]_\kappa^{(\alpha)} := \prod_{j=1}^N \Big ( c - {1 \over \alpha}
(j-1) \Big )_{\kappa_j}$.
We also have
\begin{equation} \label{l-gen4}
\Big ( \prod(1-z) \Big )^{-(a+q)} {}_0^{}{\cal F}_0^{(\alpha)}
(-x;{z \over 1 - z}) =
\sum_{\kappa}  L_\kappa^a(x;\alpha)  C_\kappa^{(\alpha)}(z)
\end{equation}
where $ \prod(1-z) := \prod_{j=1}^N (1 - z_j)$.
}

\vspace{.2cm}
\noindent
{\it Proof} \quad
In each case we need to establish that $ L_\kappa^a(x;\alpha)$ as defined by
the generating function is an eigenfunction of the operator (\ref{elpoly}), 
which in terms of the notation (\ref{defs.1}) reads
\begin{equation} \label{off.0}
\tilde{H}^{(L)} = D_1 + (a+1) E_0 - E_1,
\end{equation}
with
eigenvalue $-|\kappa|$ and has an expansion in terms of Jack polynomials
with highest weight term $(-1)^{|\kappa|} C_\kappa^{(\alpha)}(x)/|\kappa|!
C_\kappa^{\alpha}(1^N)$. The proof of the first requirement relies on
the identities
\begin{subeqnarray}
\Big ( D_1^{(x)} + (a+1)E_0^{(x)} \Big )  {}_0^{}{\cal F}_1^{(\alpha)}(a + q;
x;z) = p_1(z)  {}_0^{}{\cal F}_1^{(\alpha)}(a + q;
x;z) \slabel{goo.1} \\
\Big ( D_1^{(x)} - E_2^{(y)} \Big ) {}_0^{}{\cal F}_0^{(\alpha)}(x;y) =
(q-1)p_1(y) {}_0^{}{\cal F}_0^{(\alpha)}(x;y) \slabel{goo.2}
\end{subeqnarray}
with (\ref{goo.1}) being established in the Appendix, and (\ref{goo.2})
simply a rewrite of (\ref{mi.4}) with $y\rightarrow y/2$.

First consider (\ref{l-gen3}). We have
\begin{equation}
E_1^{(z)} {}_0^{}{\cal F}_1^{(\alpha)}(a + q;
x;-z) e^{p_1(z)} = e^{p_1(z)} E_1^{(z)} {}_0^{}{\cal F}_1^{(\alpha)}(a + q;
x;-z) + p_1(z)e^{p_1(z)} {}_0^{}{\cal F}_1^{(\alpha)}(a + q;
x;-z). \label{goo.3}
\end{equation}
Using (\ref{goo.1}) and the fact that $E_1^{(z)}$ is an eigenoperator of 
the Jack polynomials so that its action on ${}_0^{}{\cal F}_1^{(\alpha)}
(x;-z)$ is the same as the action
of $E_1^{(y)}$, the r.h.s.~of (\ref{goo.3}) can be rewritten as
\begin{equation} \label{goo.4}
\Big ( E_1^{(x)} - D_1^{(x)} - (a+1) E_0^{(x)} \Big )
{}_0^{}{\cal F}_1^{(\alpha)}(a + q;
x;-z) e^{p_1(z)}.
\end{equation}
Substituting the generating function (\ref{l-gen3}) in the l.h.s.~of 
(\ref{goo.3}) and computing the action of $E_1^{(z)}$, and comparing 
coefficients of $C_\kappa^{(\alpha)}(z)$ with (\ref{goo.4}) 
after also substituting the generating function (\ref{l-gen3})
establishes the eigenvalue equation. The expansion of $L_\kappa^{(\alpha)}(x;
\alpha)$ in terms of Jack polynomials as deduced from (\ref{l-gen3}) 
is given by
the Proposition 4.3. Its highest weight term is $(-1)^{|\kappa|}
C_\kappa^{(\alpha)}(x)/C_\kappa^{(\alpha)}(1^N)$ as required.

Now consider (\ref{l-gen4}). Setting $y_j := z_j/(1-z_j)$ this is 
equivalent to
\begin{equation}\label{fire.1}
\sum_{\mu} L_{\mu}^a(x;\al)\ca_{\mu}(\frac{y}{1+y})
= \Big ( \prod(1+y) \Big )^{a+q} {}_0^{}F_0^{(\alpha)}(-x;y)
\end{equation}
To verify the eigenvalue equation, note that $E_1^{(y)} + E_2^{(y)}$ is an
eigenoperator of $C_\mu^{(\alpha)}(y/(1+y))$ with eigenvalue $|\mu|$,
and
\begin{equation}
(E_1^{(y)} + E_2^{(y)})  \Big ( \prod(1+y) \Big )^{a+q} =
(a+q) p_1(y)  \Big ( \prod(1+y) \Big )^{a+q}.
\end{equation}
Thus, applying $E_1^{(y)} + E_2^{(y)}$ to the generating function 
(\ref{fire.1}) we have 
\begin{eqnarray}\lefteqn{
\sum_{\mu}|\mu| L_{\mu}^a(x;\alpha)\ca_{\mu}(\frac{y}{1+y})}
\nonumber \\
&= & E_1^{(y)} + E_2^{(y)})  \Big ( \prod(1+y) \Big )^{a+q}
{}_0^{}{\cal F}_0^{(\alpha)}(-x;y) \nonumber \\
& = & \Big ( \prod(1+y) \Big )^{a+q}
\Big ( E_1^{(y)} + E_2^{(y)} + (a+q)p_1(y) \Big ) 
{}_0{\cal F}_0^{(\alpha)}(-x;y)
 \nonumber \\
& = & \Big ( \prod(1+y) \Big )^{a+q}
\Big ( E_1^{(x)} - D_1^{(x)} + (a+1)p_1(y) \Big ) 
{}_0^{}{\cal F}_0^{(\alpha)}(-x;y)
 \nonumber \\
& = & \Big ( \prod(1+y) \Big )^{a+q}
\Big ( E_1^{(x)} - D_1^{(x)} - (a+1)E_0^{(x)} \Big ) 
{}_0^{}{\cal F}_0^{(\alpha)}(-x;y) \label{fire.2}
\end{eqnarray}
where to obtain the third equality we have used (\ref{goo.2}) 
while the final equality follows from (\ref{id.o}). The factor 
involving $ \prod(1+y)$ in the final equality can
be commuted in front of the operators, since they act only on the
$x$-variables. Use of the generating function (\ref{fire.1}) and comparison
of the coefficient of $C_\kappa^{(\alpha)}(y/(1+y))$ in (\ref{fire.2})
establishes the eigenvalue equation. The highest weight term in 
the Jack polynomial expansion of 
$ L_{\kappa}^a(x;\alpha)$ as defined by (\ref{l-gen4}) is the same 
as the highest
weight coefficient of $C_\kappa^{(\alpha)}(z)$ in the expansion of
$$
 \Big ( \prod(1-z) \Big )^{-(a+q)} \sum_{\mu \atop |\mu| \le |\kappa|}
{(-1)^{|\mu|} \over |\mu|!} C_\mu^{(\alpha)}(x)
{C_\mu^{(\alpha)}(z/(1-z)) \over  C_\mu^{(\alpha)}(1^N)}
$$
Consideration of the form of the expansion of $C_\mu^{(\alpha)}(z/(1-z))$
and $ \Big ( \prod(1-z) \Big )^{-(a+q)} $ in terms of 
$\{C_\si^{(\alpha)}(z)\}_\si$
shows that the highest weight coefficient of $C_\kappa^{(\alpha)}(z)$
in the above expression
is $(-1)^{|\kappa|} C_\kappa^{(\alpha)}(x)/|\kappa|! C_\kappa^{(\alpha)}(1^N)
$. Comparison with the coefficient of $C_\kappa^{(\alpha)}(z)$ on the
r.h.s.~of (\ref{l-gen4}) shows that $L_\kappa^{(\alpha)}(x;\alpha)$ 
has the required
highest weight term for its expansion in terms of Jack polynomials.

\vspace{.2cm}
In Proposition 4.1 the Laguerre polynomial is given by generating
functions involving ${}_0{\cal F}^{(\alpha)}_0$ and
${}_0{\cal F}^{(\alpha)}_1$. It is also possible to give a generating
function involving ${}_1{\cal F}^{(\alpha)}_1$ (recall (\ref{recordar})) 
which includes both these generating functions as limiting cases.

\vspace{.2cm}
\noindent
{\bf Proposition 4.2} \quad {\it We have

\begin{equation} \label{off.1}
\prod(1-z)^{-c-q}\;{}_1^{}{\cal F}_1^{(\alpha)}\left(c+q;a+q;-x;\frac{z}{1-z}
\right) = \sum_{\la}\frac{[c+q]^{(\al)}_{\la}}{[a+q]^{(\al)}_{\la}}
\;L^a_{\la}(x;\al)\,\ca_{\la}(z).
\end{equation}
}

\vspace{.2cm}
\noindent
{\it Proof} \quad
The derivation closely follows that of (\ref{l-gen4}) above, with 
(\ref{goo.2}) being replaced by the formula
\begin{equation} \label{tulum}
\left(D_1^{(x)} + (a +\frac{1-N}{\al})E_0^{(x)}\right)
{}_1^{}{\cal F}_1^{(\alpha)}(c;a;x;y) = \left(E_2^{(y)} +c p_1(y) \right)
{}_1^{}{\cal F}_1^{(\alpha)}(c;a;x;y)
\end{equation}
which is established in the Appendix.

\vspace{.2cm}
To derive the generating function (\ref{l-gen3}) from (\ref{off.1}) 
replace $z$ by $z/c$
and take the limit $c \to \infty$ using the facts that
$$
\lim_{c \to \infty} {}_1^{}{\cal F}_1^{(\alpha)}(c+q;a+q;-x;{z/c \over 1 - z/c})
= {}_0^{}{\cal F}_1^{(\alpha)}(a+q;-x;z), 
$$
$$
\lim_{c \to \infty} \prod(1-z/c)^{-c-q} = e^{p_1(z)}, \quad
\lim_{c \to \infty}[c+q]_\kappa^{(\alpha)} C_\kappa^{(\alpha)}(z/c) =
 C_\kappa^{(\alpha)}(z).
$$
The generating function (\ref{l-gen4}) follows from (\ref{off.1}) 
by setting $c=a$.

{}From the generating function (\ref{l-gen3}) it is possible to deduce the 
higher-dimensional analogue of the series expansion (\ref{lag-1v}).

\vspace{.2cm}
\noindent
{\bf Proposition 4.3} \quad
{\it We have
\begin{subeqnarray}\label{offf}
L_\kappa^a(x;\alpha) &=&
{[a+q]_\kappa^{(\alpha)}\over |\kappa|!} \sum_{\sigma\subseteq\ka}
 \left (
{\kappa \atop \sigma} \right ) {(-1)^{|\sigma|} C_\sigma^{(\alpha)}(x)
\over [a+q]_\sigma^{(\alpha)}C_\sigma^{(\alpha)}(1^N)} \slabel{off.2}\\
C_\kappa^{(\alpha)}(x) &=& [a+q]_\kappa^{(\alpha)}
C_\kappa^{(\alpha)}(1^N) \sum_{\sigma\subseteq\ka}(-1)^{|\si|}
 \left ( {\kappa \atop \sigma} \right )
{|\sigma|! L_\sigma^a(x;\alpha) \over [a+q]_\sigma^{(\alpha)}}
\slabel{off.3}
\end{subeqnarray}
}

\vspace{.2cm}
\noindent
{\it Proof} \quad The formula (\ref{off.2}) follows from the generating 
function
formula (\ref{l-gen3}) by applying the identity \cite{lass90a,kaneko93a}
\begin{equation}\label{id.7}
e^{p_1(z)}\,\ca_{\la}(z) = \sum_{\mu}\bin{\mu}{\la}\frac{|\la|!}
{|\mu|!}\ca_{\mu}(z)
\end{equation}
on the l.h.s.~and equating coefficients of $C_\kappa^{(\alpha)}(z)$.
The formula (\ref{off.3}) follows from (\ref{l-gen3}) by 
multiplying both sides by
$e^{-p_1(z)}$, using the identity (\ref{id.7}) (with $z$ replaced by $-z$)
on the r.h.s., and equating coefficients of $C_\kappa^{(\alpha)}(z)$.

A simple consequence of the generating function (\ref{l-gen4}) 
is the analogue of
the formula of Proposition 3.6, which is derived by following the steps
of the proof of that formula.

\vspace{.2cm}
\noindent
{\bf Proposition 4.4} \quad {\it We have
$$
L_k^{N(a+q) - 1} (p_1(y)) = \sum_{|\kappa| = k}
L_\kappa^a(y;\alpha) C_\kappa^{(\alpha)}(1^N).
$$
}

\vspace{.2cm}
\noindent
Since each $L_\kappa^a(y;\alpha)$ is an eigenfunction of (\ref{off.0}) with
eigenvalue $-|\kappa|$, it follows that $L_k^{N(a+q) - 1} (p_1(y))$
is also an eigenfunction of (\ref{off.0}) with eigenvalue $-k$ (this feature 
can be checked directly).

As our final result of this section, we note that the proof of (\ref{l-gen3})
can be generalized to give a family of eigenoperators for
$L_\kappa^a(y;\alpha)$, together with the corresponding eigenvalues. These
operators are analogues of the operators of Proposition 3.2 for the
generalized Hermite polynomials, and are derived in the same way.

\vspace{.2cm}
\noindent
{\bf Proposition 4.5} \quad {\it Let
\begin{eqnarray*}
\tilde{H}^{(L)}{(y)} := \left(D_N^{p\,(y)} - \left[D_1^{(y)}+(a+1)E_0^{(y)},
D_N^{p\,(y)} \right] +\cdots \right. \hspace{4cm}\\
\left.+\frac{(-1)^n}{n!}\left[D_1^{(y)}+(a+1)E_0^{(y)},\left[\cdots
\left[D_1^{(y)}+(a+1)E_0^{(y)},D_N^{p\,(y)}\right]\cdots\right]\right]
\right)
\end{eqnarray*}
where $D_N^p$ is the operator introduced in Section 3.1.
We have that $L_\kappa^a(y;\alpha)$ is an
eigenfunction of $\tilde{H}^{(L)}{(y)}$ for each $p=1,\dots,N$, with eigenvalue
$e_p(\kappa;\alpha)$ given by the coefficient of $X^{N-p}$ 
in (\ref{gev}).
}

\subsection{Recurrence and differentiation formulas}

The classical Laguerre polynomials satisfy the recurrence relations
\begin{subeqnarray}
xL_n^a(x)& =& (2n+a+1)L_n^a(x) - (n+1)L_{n+1}^a(x) - (n+a)L_{n-1}^a(x) 
\slabel{l.1}\\
L_n^{a+1}(x)& =& \sum_{m=0}^n L_m^a(x)\slabel{l.2}\\
L_n^a(x)& =& L_n^{a+1}(x) -  L_n^{a+1}(x)\slabel{l.3}
\end{subeqnarray}
and the differentiation formulas
\begin{subeqnarray}
{d \over dx} L_n^a(x)& =& - L_{n-1}^{a+1}(x)  \slabel{l.4}\\
x{d \over dx} L_n^a(x)& =& nL_n^a(x) - (n+a) L_{n-1}^a(x). \slabel{l.5}
\end{subeqnarray}
The generalized Laguerre polynomials satisfy higher-dimensional analogues
 of these formulas. Let us first consider (\ref{l.5}).

\vspace{.2cm}
\noindent
{\bf Proposition 4.6} \quad {\it We have
$$
E_1^{(x)}\,L_{\kappa}^a(x;\alpha) = |\kappa|L_{\kappa}^a(x;\alpha)
-{1 \over |\kappa|} \sum_i\bin{\kappa}
{\kappa_{(i)}}\left(\kappa_i + a +\frac{N-i}{\alpha}\right)
L^a_{\kappa_{(i)}}(x;\alpha)
$$
}

\vspace{2mm}
\noindent {\it Proof} \quad From the generating function (\ref{fire.1}) 
and the fact
that $E_1^{(x)}$ is an eigenoperator of $C_\kappa^{(\alpha)}(x)$ we have
\begin{eqnarray}\lefteqn{
E_1^{(x)}\sum_{\mu}L_{\mu}^a(x;\alpha)C_{\mu}^{(\alpha)}
(\frac{y}{1+y})} \nonumber \\
& = & \prod(1+y)^{a+q} E_1^{(y)} {}_0^{}{\cal F}^{(\alpha)}_0(-x;y)
\nonumber \\& = &
\left(E_1^{(y)}-(a+q)p_1(y/(1+y)) \right)  \prod(1+y)^{a+q}
 {}_0^{}{\cal F}^{(\alpha)}_0(-x;y) \nonumber \\
& =& \left(E_1^{(y)} -(a+q)p_1(y/(1+y)) \right)
\sum_{\mu}L_{\mu}^a(x;\alpha)C_{\mu}^{(\alpha)}
(\frac{y}{1+y}) \label{guff.1}
\end{eqnarray}
But with $z_j := y_j/(1+y_j)$,
\begin{eqnarray*}
E_1^{(y)} C_{\mu}^{(\alpha)}(\frac{y}{1+y}) &=& (E_1^{(z)} - E_2^{(z)})
 C_{\mu}^{(\alpha)}(z) \\
&=& |\mu| C_{\mu}^{(\alpha)}(z) - \frac{1}{1+|\mu|}\sum_{i=1}^N
\bin{\mu^{(i)}}{\mu}\left(\mu_i-\frac{i-1}{\alpha}
\right)\,C_{\mu^{(i)}}^{(\alpha)}(z)
\end{eqnarray*}
where the second equality uses (\ref{id.3}). Substituting 
this expression in the r.h.s.~of (\ref{guff.1}), and using 
(\ref{id.oo}) to simplify the remaining term on the
r.h.s.~of (\ref{guff.1}) gives
\begin{eqnarray*}\lefteqn{
E_1^{(x)}\sum_{\mu}L_{\mu}^a(x;\al)C_{\mu}^{(\alpha)}
(\frac{y}{1+y}) = \sum_{\mu}L_{\mu}^a(x;\alpha)
\left\{ |\mu| C_{\mu}^{(\alpha)}(\frac{y}{1+y}) \right.} \\ & &
\left. - \frac{1}{1+|\mu|}\sum_{i=1}^N
\bin{\mu^{(i)}}{\mu}\left(\mu_i+1+a +\frac{N-i}{\alpha}
\right)\,C_{\mu^{(i)}}^{(\alpha)}(\frac{y}{1+y}) \right\}
\end{eqnarray*}
The result follows by equating coefficients of $C_{\mu}^{(\alpha)}
(\frac{y}{1+y})$.

\vspace{.2cm}
Next we will derive the analogue of (\ref{l.1}).

\vspace{.2cm}
\noindent
{\bf Proposition 4.7} \quad {\it We have
\begin{eqnarray*}
p_1(x)\,L^a_{\kappa}(x;\alpha) = \left(2|\kappa| +N(a+q)\right)L^a_{\kappa}
(x;\alpha) + {1 \over |\kappa|}
\sum_{i=1}^N \bin{\kappa}{\kappa_{(i)}}
\left(\kappa_i+a+\frac{N-i}{\alpha}\right)
L^a_{\kappa_{(i)}}(x;\alpha) \\
-(|\kappa| + 1) \alpha\sum_i\bin{\kappa^{(i)}}{\kappa}
\frac{j_{\kappa}}{j_{\kappa^{(i)}}}
\left(N-i+1+\alpha\kappa_i\right)\,L^a_{\kappa^{(i)}}(x;\alpha)
\end{eqnarray*}
}

\vspace{.2cm}
\noindent  {\it Proof} \quad From the generating function (\ref{fire.1})
\begin{eqnarray}
\lefteqn{\sum_{\mu}p_1(x) L_{\mu}^a(x;\alpha)C_{\mu}^{(\alpha)}
(\frac{y}{1+y}) = p_1(x)  \prod(1+y)^{a+q} {}_0^{}{\cal F}^{(\alpha)}_0(-x;y)}
\nonumber \\
&& = -  \prod(1+y)^{a+q} E_0^{(y)}  {}_0^{}{\cal F}^{(\alpha)}_0(-x;y) 
\nonumber \\
&&=  -E_0^{(y)}  \prod(1+y)^{a+q}  {}_0^{}{\cal F}^{(\alpha)}_0(-x;y)
+  {}_0^{}{\cal F}^{(\alpha)}_0(-x;y) E_0^{(y)}
 \prod(1+y)^{a+q} \nonumber \\
&& = \Big (  -E_0^{(y)} + (a+q) p_1(1/(1+y)) \Big ) 
 \prod(1+y)^{a+q}  {}_0^{}{\cal F}^{(\alpha)}_0(-x;y)
 \nonumber \\ && =
\Big (  -E_0^{(y)} + (a+q) p_1(1/(1+y)) \Big )
\sum_{\mu}L_{\mu}^a(x;\alpha)C_{\mu}^{(\alpha)}
(\frac{y}{1+y}) \label{off.7}
\end{eqnarray}
The task is now to write $E_0^{(y)} \, C_{\mu}^{(\alpha)}
(\frac{y}{1+y})$ and $ p_1(1/(1+y)) C_{\mu}^{(\alpha)}
(\frac{y}{1+y})$ as a series in \linebreak $\{  C_{\kappa}^{(\alpha)}
(\frac{y}{1+y})\}_\kappa$. To do this let $z_j = y_j/(1+y_j)$ so
that
$$
E_0^{(y)} = \sum_{j=1}^N (1 - z_j)^2 {\partial \over
\partial z_j} = E_0^{(z)} - 2E_1^{(z)} + E_2^{(z)} \quad
{\rm and} \quad p_1(1/(1+y)) = p_1(1-z) = N - p_1(z).
$$
We then have
\begin{eqnarray}
E_0^{(y)} \, C_{\mu}^{(\alpha)} 
(\frac{y}{1+y})& =& \Big (  E_0^{(z)} - 2E_1^{(z)} + E_2^{(z)}
\Big )  C_{\mu}^{(\alpha)}(z) \nonumber \\
& = &  \sum_{i=1}^N \bin{\mu}{\mu_{(i)}}\,
\frac{C^{(\alpha)}_{\mu}(1^N)}{C^{(\alpha)}_{\mu_{(i)}}
(1^N)} C^{(\alpha)}_{\mu_{(i)}}(\frac{y}{1+y})
-2 |\mu|  C^{(\alpha)}_{\mu}(\frac{y}{1+y}) \nonumber \\
& & + \frac{1}{1+|\mu|}\sum_{i=1}^N
\bin{\mu^{(i)}}{\mu}\left(\mu_i-\frac{i-1}{\alpha}
\right)\,C_{\mu^{(i)}}(\frac{y}{1+y}) \label{off.5}
\end{eqnarray}
and
\begin{eqnarray}
 p_1(1/(1+y)) C_{\mu}^{(\alpha)}
(\frac{y}{1+y}) & = & \Big (  N - p_1(z) \Big )  C_{\mu}^{(\alpha)}
(z) \nonumber \\ \label{off.6}
& = & N  C_{\mu}^{(\alpha)}
(\frac{y}{1+y}) - \frac{1}{1+|\mu|}\sum_{i=1}^N
\bin{\mu^{(i)}}{\mu} C_{\kappa^{(i)}}^{(\alpha)}(\frac{y}{1+y})
\end{eqnarray}
where to obtain (\ref{off.5}) we have used (\ref{id.1}), (\ref{id.2}) 
and (\ref{id.3}), while
to obtain (\ref{off.6}) we have used (\ref{id.oo}). Substituting 
(\ref{off.5}) and (\ref{off.6}) in the r.h.s.~of (\ref{off.7}), 
equating coefficients of $ C_{\kappa^{(i)}}^{(\alpha)}
(\frac{y}{1+y})$ with the l.h.s.~of (\ref{off.7}), and use of 
(\ref{ci.1}) to
rewrite $C^{(\alpha)}_{\mu}(1^N)/C^{(\alpha)}_{\mu_{(i)}}
(1^N)$ gives the stated result. 

The generalizations of (\ref{l.2}) and (\ref{l.3}) are given by the
following result.

\vspace{2mm}\noindent
{\bf Proposition 4.8} \quad {\it We have 
\begin{eqnarray}
L_{\ka}^{a-1}(x;\al) &=& \sum_{r=0}^{\mbox{\scriptsize min}(N,|\kappa|)}
(-\alpha)^r \sum_{\stackrel{\si}{\ka/\si\mbox
{\scriptsize\ a vertical $r$-strip}}} \frac{|\sigma|!}
{|\kappa|!} \psi_{\ka/\si}(\al)L^a_{\si}(x;\al) \label{ansett.1}\\
L_{\ka}^{a+1/\al}(x;\al) &=& \sum_{r=0}^{|\kappa|} \al^{-r}
\sum_{\stackrel{\si}{\ka/\si\mbox
{\scriptsize\ a horizontal $r$-strip}}} \frac{|\sigma|!}
{|\kappa|!} \phi_{\ka/\si}(\al)L^a_{\si}(x;\al) \label{ansett.2}
\end{eqnarray}
where
\begin{subeqnarray}
\psi_{\ka/\si}(\al)=\prod_{s\in R_{\ka/\si}\cap \ka}
\frac{h_{\ka}^*(s)}{h^{\ka}_*(s)}
\prod_{s\in R_{\ka/\si}\cap \si}
\frac{h^{\si}_*(s)}{h_{\si}^*(s)}\prod_{s\in\ka}
h_*^{\ka}(s) \prod_{s\in\si}(h_*^{\si}(s))^{-1} \slabel{off.8}\\
\phi_{\ka/\si}(\al)=
\prod_{s\in C_{\ka/\si}\cap \ka}
\frac{h^{\ka}_*(s)}{h_{\ka}^*(s)}
\prod_{s\in C_{\ka/\si}\cap \si}
\frac{h_{\si}^*(s)}{h^{\si}_*(s)}\prod_{s\in\ka}
h^*_{\ka}(s) \prod_{s\in\si}(h^*_{\si}(s))^{-1}
\end{subeqnarray}
Here $R_{\ka/\si}$ denotes the union of all rows which
intersect $\ka - \si$, 
$C_{\ka/\si}$ denotes the union of all columns which
intersect $\ka - \si$, and $h_{\ka}^*(s)$, $h^{\ka}_*(s)$
etc. are given by (\ref{mi.3}).
}

\vspace{2mm}\noindent{\it Proof}\quad
First consider (\ref{ansett.1}). From the generating function
(\ref{l-gen4}) we see that
\begin{equation} \label{sea.2}
\sum_{\mu} L^{a-1}_{\mu}(x;\alpha)C^{(\alpha)}_{\mu}(z) = 
\prod_{j=1}^N(1-z_j) \sum_{\si} L^{a}_{\si}(x;\alpha) C^{(\alpha)}_{\si}(z)
\end{equation}
Using\cite{stan89a}
$$
\prod_{j=1}^N(1-z_j) = \sum_{r=0}^N \frac{(-1)^r}{r!}\,
J_{(1^r)}^{(\alpha)}(z)
$$
the Pieri formula\cite{stan89a}
\begin{equation}\label{pieri.1}
J^{(\al)}_{1^r}\,J^{(\al)}_{\si} = r!\,\sum_{\stackrel{\ka}{\ka/\si\mbox
{\scriptsize\ a vertical $r$-strip}}} \frac{j_{\si}}{j_{\ka}}
\psi_{\ka/\si}(\al) J^{(\al)}_{\ka},
\end{equation}
and the relationship (\ref{farfel}), the r.h.s. of (\ref{sea.2})
can be rewritten as
\begin{equation}\label{sea.3}
\sum_{r=0}^N\sum_{\stackrel{\kappa,\si}{\ka/\si\mbox
{\scriptsize\ a vertical $r$-strip}}} (-1)^r \alpha^{|\si|}
|\si|!\ L^a_{\si}(x;\alpha)\psi_{\ka/\si}(\al)\frac{\alpha^{-|\ka|}}
{|\ka|!}\,C^{(\alpha)}_{\kappa}(z)
\end{equation}
The result now follows by comparing the coefficients of
$C^{(\alpha)}_{\kappa}(z)$ on the l.h.s. of (\ref{sea.2}) with
that in (\ref{sea.3}). 

Now consider (\ref{ansett.2}). Using the formula \cite{stan89a}
$$
\prod_{j=1}^N (1 - z_j)^{-1/\alpha} = \sum_{r=0}^\infty 
{J_{(r)}^{(\alpha)}(z) \over \alpha^r r!}
$$
and the Pieri formula (\ref{pieri.1}) in conjugate form
$$
J_{(r)}^{(\alpha)}(z) ,J_{\si}^{(\alpha)}(z) = r! \alpha^r
\,\sum_{\stackrel{\ka}{\ka/\si\mbox
{\scriptsize\ a horizontal $r$-strip}}} \frac{j_{\si}}{j_{\ka}}
\phi_{\ka/\si}(\al) J_{\ka}^{(\alpha)}(z),
$$
together with the formula (\ref{farfel}), we have
\begin{eqnarray*}
\sum_{\mu} L_\mu^{a+1/\alpha}(x;\alpha) C^{(\alpha)}_{\mu}(z) &=& 
\prod_{j=1}^N (1 - z_j)^{-1/\alpha} \sum_{\sigma} L_\sigma^{a}(x;\alpha) 
C^{(\alpha)}_{\sigma}(z) \\
& = & \sum_{r=0}^\infty \,\sum_{\stackrel{\ka}{\ka/\si\mbox
{\scriptsize\ a horizontal $r$-strip}}}  L_\sigma^{a}(x;\alpha)
\alpha^{|\sigma| - |\kappa|} 
{|\sigma|!\over |\kappa|!} \phi_{\ka/\si}(\al)  C^{(\alpha)
}_{\sigma}(z).
\end{eqnarray*}
The identity (\ref{ansett.2}) follows by equating coefficients of
$ C^{(\alpha)}_{\sigma}(z)$.

\vspace{.2cm}
The analogue of (\ref{l.4}) takes the form

\vspace{2mm}\noindent
{\bf Proposition 4.9} {\it
$$
E_0^{(x)} L^a_{\ka}(x;\al) = \sum_{r=1}^{\mbox{\scriptsize min}(N,|\kappa|)}
r\alpha^{-r}
\sum_{\stackrel{\si}{\ka/\si\mbox{\scriptsize\ a vertical $r$-strip}}}
\frac{|\sigma|!}{|\kappa|!} \psi_{\ka/\si}(\al)L^{a+1}_{\si}(x;\al)
$$
where $\psi_{\ka/\si}(\al)$ is given by (\ref{off.8}).
}

\vspace{2mm}\noindent{\it Proof} \quad
Applying $E_0^{(x)}$ to the generating function (\ref{l-gen4}), we have
\begin{eqnarray}
E_0^{(x)} \sum_{\kappa}L^a_{\kappa}(x;\alpha)C^{(\alpha)}_{\kappa}(z)
&=& \prod_j(1-z_j)^{-a-q}E_0^{(x)}{}_0{\cal F}_0\left(-x,\frac{z}{1-z}
\right) \nonumber\\
&=& -p_1\left(\frac{z}{1-z}\right) \prod_j(1-z_j)^{-a-q} {}_0{\cal F}_0
\left(-x,\frac{z}{1-z} \right) \nonumber\\
&=& -p_1\left(\frac{z}{1-z}\right) \prod_j(1-z_j) \sum_{\sigma}
L^{a+1}_{\sigma}(x;\alpha) C^{(\alpha)}_{\sigma}(z)\label{pocari.1}
\end{eqnarray}
Certainly
$$
p_1\left(\frac{z}{1-z}\right) \prod_j(1-z_j) = 
\sum_k z_k \prod_{p\neq k}(1-z_p).
$$
If we differentiate w.r.t. $t$ the identity
$$
\prod_{i=1}^N(1-z_it) = \sum_{r=0}^N\frac{(-1)^r}{r!}
J^{(\al)}_{1^r}(z) t^r
$$
giving
$$
\sum_{r=1}^N \frac{(-1)^r}{(r-1)!}J^{(\al)}_{1^r}(z) t^{r-1}
= -\sum_k z_k \prod_{p\neq k}(1-z_p t)
$$
and set $t=1$, we obtain
\begin{equation} \label{pocari.2}
-p_1\left(\frac{z}{1-z}\right) \prod_j(1-z_j) = 
\sum_{r=1}^N \frac{(-1)^r}{(r-1)!}J^{(\al)}_{1^r}(z)
\end{equation}
Inserting (\ref{pocari.2}) back into (\ref{pocari.1}) gives, after
some manipulation
$$
E_0^{(x)} \sum_{\kappa}L^a_{\kappa}(x;\alpha)C^{(\alpha)}_{\kappa}(z)
= \sum_{r=1}^{N} r\alpha^{-r}
\sum_{\stackrel{\kappa,\si}{\ka/\si\mbox{\scriptsize\ a vertical $r$-strip}}}
\frac{|\sigma|!}{|\kappa|!} \psi_{\ka/\si}(\al)L^{a+1}_{\si}(x;\al)
C^{(\alpha)}_{\kappa}(z)
$$
which yields the result upon comparison of the coefficients
of $C^{(\alpha)}_{\kappa}(z)$.
 
\subsection{Integration formulas}

The classical Laguerre polynomial obeys integration formulas analogous
to the integration formulas (\ref{hint.0}) for the 
classical Hermite polynomial:
\begin{subeqnarray}
\int_0^\infty y^a e^{-y} \Big ( L_k^a(y) \Big )^2\, dy  =
{\Gamma (a+1+k) \over k! }&& \slabel{lint.1}\\
{k! \over \Gamma(a+1)}e^x
\int_0^\infty y^a e^{-y} {}_0F_1(a+1;-xy) L_k^a(y) \, dy & = & x^k 
\slabel{lint.2}\\
{e^x \over k! \Gamma(a+1)} \int_0^\infty y^a e^{-y} {}_0F_1(a+1;-xy)
y^k  \, dy & = & L_k(x) \slabel{lint.3}
\end{subeqnarray}
The higher-dimensional analogues of these formulas can be established using
the generating functions (\ref{l-gen3}), (\ref{l-gen4}) 
in much the same way as the higher-dimensional
analogues of (\ref{hint.0}) 
were established using the generating function (\ref{gfhermite}).
To present these results, we will make use of the notation
\begin{equation}
d\mu^{(L)}(y) := \prod_{j=1}^N y_j^a\ e^{-y_j} \prod_{1\le j < k \le N}
|y_k - y_j|^{2/\alpha}dy_1 \dots dy_N.
\end{equation}

\vspace{.2cm}
\noindent
{\bf Proposition 4.10} \quad {\it We have
\begin{equation} \label{tenga}
{\cal N}_\kappa^{(L)} := \int_{[0,\infty)^N} \Big (
L_\kappa^a(x;\alpha)\Big )^2 d\mu^{(L)}(x) = {\cal N}_0^{(L)}
{[a+q]_\kappa^{(\alpha)} \over C_\kappa^{(\alpha)}(1^N) |\kappa|!}
\end{equation}
where
\begin{equation}
 {\cal N}_0^{(L)} :=  \int_{[0,\infty)^N}  d\mu^{(L)}(x) =
\alpha^{(1-N-(N-1)^2/\alpha)} \prod_{j=0}^{N-1}
{\Gamma(1+(j+1)/\alpha) \Gamma(a+1+j/\alpha) \over \Gamma(1+1/\alpha)}.
\end{equation}
}

\vspace{.2cm}
\noindent
{\it Proof} \quad Multiplication of both sides of the generating function
(\ref{l-gen4}) by $L_{\kappa}^a(x;\alpha)$ and integration with respect to
$ d\mu^{(L)}(x)$ gives, upon using the orthogonality of
$\{L_{\kappa}^a \}_\kappa $ with respect to the inner product (\ref{il}),
$$
\prod (1-z)^{-(a+q)} \int_{[0,\infty)^N}
 {}_0^{}{\cal F}^{(\alpha)}_0(-x;{z \over 1 - z}) L_{\kappa}^a(x;\alpha)
\, d\mu^{(L)}(x) = {\cal N}_\kappa^{(L)} C_\kappa^{(\alpha)}(z).
$$
Setting $z_1 = \dots = z_N = c$, using (\ref{dia.1}) and changing variables
$cx_j/(1-c) =: y_j$ this reads
$$
c^{-N(a+q)}  \int_{[0,\infty)^N} e^{-p_1(y)/c}\prod_{j=1}^N y_j^a \,
L_{\kappa}^a\left(\frac{(1-c)}{c}y_j\right) 
\prod_{1\leq j<k\leq N}|y_k - y_j|^{2/\alpha}\,
\prod_{j=1}^N dy_j =  {\cal N}_\kappa^{(L)}  C_\kappa^{(\alpha)}(c^N).
$$
The stated result follows by choosing $c=1$ and noting from (\ref{off.2})
that
\begin{equation}
 L_{\kappa}^a(0;\alpha) = [a+q]_\kappa^{(\alpha)} / |\kappa|!.
\end{equation}

\vspace{.2cm}
The analogues of the formulas (\ref{lint.2}) and (\ref{lint.3}) are 
deduced from the following integration formula.

\vspace{.2cm}
\noindent
{\bf Proposition 4.11} \quad {\it We have
\begin{eqnarray}
 \int_{[0,\infty)^N} {}_0^{}{\cal F}^{(\alpha)}_1(a+q;x;-z_a)\,
 {}_0^{}{\cal F}^{(\alpha)}_1(a+q;x;-z_b) \,  d\mu^{(L)}(x) \nonumber \\
=  {\cal N}_0^{(L)} e^{-p_1(z_a)}e^{-p_1(z_b)}
 {}_0^{}{\cal F}^{(\alpha)}_1(a+q;z_a;z_b). \label{leche}
\end{eqnarray}
}

\vspace{.2cm}
\noindent
{\it Proof} \quad Substitute for $ e^{p_1(z_s)} 
{}_0^{}{\cal F}^{(\alpha)}_1(a+q;x;-z_s)$ $(s=a,b)$ using 
the generating function (\ref{l-gen3}) and integrate
with respect to $ d\mu^{(L)}(x)$
term-by-term. From the orthogonality property of $\{  L_{\kappa}^a \}_\kappa$
with respect to (\ref{il}), only the diagonal terms 
in the double sum are non-zero,
with the integral then being evaluated according to Proposition 4.10.
The resulting sum is identified with $ {}_0^{}{\cal F}^{(\alpha)}_1$
according to the definition (\ref{recordar}).

\vspace{.2cm}
\noindent
{\bf Corollary 4.1} \quad
{\it We have
\begin{subeqnarray}\label{tdd} 
 \int_{[0,\infty)^N} {}_0^{}{\cal F}^{(\alpha)}_1(a+q;x;-z_a)
 L_{\kappa}^a(x;\alpha) \,  d\mu^{(L)}(x)
= { {\cal N}_0^{(L)} \over C_\kappa^{(\alpha)}(1^N)
|\kappa|!} e^{-p_1(z_a)}C_\kappa^{(\alpha)}(z_a)
\hspace{15mm}\slabel{td.1}\\
 \int_{[0,\infty)^N} {}_0^{}{\cal F}^{(\alpha)}_1(a+q;x;-z_a)
 C_\kappa^{(\alpha)}(x)  \,  d\mu^{(L)}(x) =
 {\cal N}_0^{(L)}  C_\kappa^{(\alpha)}(1^N)
|\kappa|!  e^{-p_1(z_a)}  L_{\kappa}^a(z_a;\alpha). \qquad\slabel{td.2}
\end{subeqnarray}
}

\vspace{.2cm}
\noindent
{\it Proof} \quad The integration formula (\ref{td.1}) follows 
from (\ref{leche}) after
multiplying both sides by $ e^{p_1(z_b)}$, using the generating function
(\ref{l-gen3}) to substitute for 
$ e^{p_1(z_b)} {}_0^{}{\cal F}^{(\alpha)}_1(a+q;x;-z_b)$ and 
equating coefficients of $C_\kappa^{(\alpha)}(z_b)$ on both sides.
The integration formula (\ref{td.2}) follows from (\ref{leche}) 
after replacing $z_b$ by $-z_b$, substituting for 
$ e^{p_1(z_b)} {}_0^{}{\cal F}^{(\alpha)}_1(a+q;z_a;-z_b)$ using 
(\ref{l-gen3}) and equating coefficients of $C_\kappa^{(\alpha)}(z_b)$
on both sides.
 
Analogous to the Hermite case, the integration formula (\ref{leche}) 
can be used to derive the analogue of the classical summation formula
valid for $|t|<1$
\begin{equation}
\sum_{n=0}^\infty {n! \over (a + 1)_n}
L_n^a (x) L_n^a (y) t^n = (1 - t)^{-a-1}
\exp \Big ( - \frac{t}{1-t}(x + y) \Big )
{}_0F_1\left(a+1;\frac{xyt}{(1-t)^2}\right)
\end{equation}

\vspace{.2cm}
\noindent
{\bf Proposition 4.12} \quad {\it For $|t|<1$ we have 
\begin{eqnarray*}
G^{(L)}(x,y;t)
:= \sum_{\kappa} \frac{L_{\kappa}(x;\alpha)L_{\kappa}(y;\alpha)}
{ {\cal N}_{\kappa}^{(L)}} t^{|\kappa|} \hspace{8cm}\\
=\frac{1}{{\cal N}_0^{(L)}} (1 - t)^{-N(a+q)}
\exp \Big ( - {t \over 1 - t}(p_1(x) + p_1(y) ) \Big )
{}_0^{}{\cal F}^{(\alpha)}_1\left( a+q;{y \over 1 - t};{tx \over 1 - t}
\right )
\end{eqnarray*}
}

\vspace{.2cm}
\noindent
{\it Proof} \quad This follows by following the procedure used in the Hermite
case, Proposition 3.9.

\vspace{.2cm}
Notice that in the special case $x=0$ the above summation reduces to
elementary functions, giving
\begin{equation} \label{cat.2}
G^{(L)}(0,y;t) = \frac{1}{{\cal N}_0^{(L)}} (1 - t)^{-N(a+q)}
\exp \Big ( - {t \over 1 - t} p_1(y) \Big ).
\end{equation}
Interpretation of this result in terms of an
explicit solution of the Fokker-Planck equation (\ref{fp-eqn})
with $W$ given by (\ref{lagpot}) will be discussed in the next
section.

Finally, we note (prompted by M.~Lassalle) that (\ref{goo.1}) 
implies the identity
\begin{equation} \label{moops.1}
{(-1)^{|\kappa|} \over |\kappa|! C_\kappa^{(\alpha)}(1^N)}
\exp \Big ( - D^{(x)}_1 - (a+1) E_0^{(x)} \Big ) C_\kappa^{(\alpha)}(x) =
L_\kappa^a(x;\alpha)
\end{equation}
(the derivation parallels that of (\ref{china.1})). Analogous 
to (\ref{china.2}), comparison with (\ref{td.2}) gives that 
for any symmetric analytic function $f(x)$,
\begin{equation}
{e^{p_1(z)} \over {\cal N}_0^{(L)}} \int_{[0,\infty)^N}
{}_0^{}{\cal F}^{(\alpha)}_1(a+q;x;-z) f(-x) d\mu^{(L)}(x) 
= \exp \Big ( - D_1^{(z)} - (a+1) E_0^{(z)} \Big ) f(z)
\end{equation}
and hence by a similar argument as before, if
\begin{equation} \label{Hank}
F(z) = {e^{p_1(z)} \over {\cal N}_0^{(L)}} \int_{[0,\infty)^N}
{}_0^{}{\cal F}^{(\alpha)}_1(a+q;x;-z) f(-x) d\mu^{(L)}(x)
\end{equation}
then
\begin{equation}\label{moops.2}
f(z) = {e^{-p_1(z)} \over {\cal N}_0^{(L)}} \int_{[0,\infty)^N}
{}_0^{}{\cal F}^{(\alpha)}_1(a+q;x;z) F(x) d\mu^{(L)}(x)
\end{equation}
(in the case $\alpha = \infty$ (\ref{Hank}) corresponds to the 
Hankel transform).

\setcounter{equation}{0}
\section{Applications}
\subsection{The ground state global density}

As noted in the Introduction, the ground states of the Schr\"odinger
operators (\ref{schrops1}) are, up to normalization, of the form
$e^{-\beta W /2}$ where $W$ is given by (\ref{pots}). The ground state
density in a system of $N+1$ particles, $\rho_{N+1}(x)$ say, is then given
by
\begin{equation}\label{p.1}
\rho_{N+1}(x) = {N+1 \over Z_{N+1}} \prod_{l=1}^N \int_I dx_l \,
e^{- \beta W(x,x_1, \dots, x_N)} 
\end{equation}
where
\begin{equation}
 Z_{N+1} := \prod_{l=1}^{N+1} \int_I dx_l \,
e^{- \beta W(x_1,x_2, \dots, x_{N+1})} 
\end{equation}
An alternative  interpretation of (\ref{p.1}) is as the density at the point
$x$ in the statistical mechanical system of $N+1$ particles with potential
energy $W$ confined to
the interval $I$, in equilibrium at inverse temperature $\beta$.

In the case $W = W^{(H)}$ as given by (\ref{hermpot}), a physical argument 
based on the interpretation of the harmonic term as an electrostatic potential
(see e.g.~\cite{brezin78}) predicts that for all $\beta$
\begin{equation}
\lim_{N \to \infty} \sqrt{{2 \over N}} \rho(\sqrt{2 N} x) =
\left \{ \begin{array}{ll} {2 \over \pi} \sqrt{ 1 - x^2}, &
|x| < 1 \\
0, & |x| \ge 1 \end{array} \right . \label{wig.h}
\end{equation}
This limit gives the so-called global density, and the result is known as
the Wigner semi-circle law. 

For $W = W^{(L)}$ the change of variables
$y_j = x_j^2$ gives
\begin{equation}
\rho_{N+1}(y) = {N+1 \over Z_{N+1}}\,e^{-\beta y/2}\,y^{\beta \mu /2} 
 \prod_{l=1}^N \int_0^{\infty} dy_l |y - y_l|^\beta \,
e^{-\beta y_l / 2} y_l^{\beta \mu /2} \prod_{1 \le j < k \le N}
|y_k - y_j|^\beta
\end{equation}
where $\mu := a' - 1/\beta$ and
\begin{equation}
 Z_{N+1} := \prod_{l=1}^{N+1} \int_0^\infty dy_l \, 
e^{-\beta y_l / 2} y_l^{\beta \mu /2} \prod_{1 \le j < k \le N+1}
|y_k - y_j|^\beta
\end{equation}
The same type of electrostatics calculation used to obtain 
(\ref{wig.h}) predicts that for all $\beta$ and $\mu$,
\begin{equation}
\lim_{N \to \infty} \rho_{N+1} (4 N y)= \left \{ \begin{array}{ll} {1 \over 2 
\pi y^{1/2}} \sqrt{ 1 - y}, &
0 < y < 1 \\
0, & y<0, \:y \ge 1 \end{array} \right . \label{wig.l}
\end{equation}

Finally, for $W = W^{(J)}$, the change of variable $\sin^2 \phi_j = y_j $
gives
\begin{equation}
\rho_{N+1}(y) = {N+1 \over Z_{N+1}} \,y^{\beta\mu_1/2}
(1-y)^{\beta\mu_2/2} \prod_{l=1}^N \int_{0}^1 dy_l \,
|y-y_l|^\beta y_l^{\beta \mu_1 / 2} (1 - y_l )^{\beta \mu_2 /2} 
\prod_{1 \le j < k \le N} |y_k - y_j|^\beta
\end{equation}
where $\mu_1 := a' - 1/\beta$, $\mu_2 := b' - 1/\beta$ and
\begin{equation}
Z_{N+1} = \prod_{l=1}^{N+1} \int_{0}^1 dy_l \,
y_l^{\beta \mu_1 / 2} (1 - y_l )^{\beta \mu_2 /2} 
\prod_{1 \le j < k \le N+1} |y_k - y_j|^\beta
\end{equation}

In this case the electrostatics calculation gives
\begin{equation}
\lim_{N \to \infty} {1 \over N} \rho_{N+1}(y)
= \left \{ \begin{array}{ll} {1 \over 
\pi } {1 \over \sqrt{y(1-y)}}&
0 < y < 1 \\
0 & y < 0, \: y > 1 \end{array} \right . \label{wig.j}
\end{equation}

In this section we will show how for $\beta $ even the density
(\ref{p.1}) in the Hermite, Laguerre and Jacobi cases is related to
eigenstates of the operator (\ref{gaugeham}) and thus the
generalized Hermite, Laguerre and Jacobi polynomials respectively
(this result is already implicit in earlier publications
\cite{forr92a,forr94b,kaneko93a}). Furthermore, 
we will show how the global density can be
evaluated by using integral representations.

\subsection{Relationship between the density and 
the generalized polynomials}

Instead of considering the density directly, we proceed as in 
\cite{forr92a} and introduce a function $f$ depending on 
the auxilary variables $t_1, \dots, t_m$:
\begin{equation}
f(t_1, \dots, t_m) := {1 \over Q_N}  \prod_{l=1}^N \int_I dy_l \,
e^{-\beta V(y_l)} \prod_{s=1}^m \prod_{l=1}^N (y_l - t_s)
\prod_{1 \le j < k \le N} |y_k - y_j|^\beta \label{llave.1}
\end{equation}
where the normalization $Q_N$ is chosen so that $f$ equals unity at
the origin.
For an appropriate choice of $I$ and $V$, (\ref{llave.1}) 
gives each of the densities
in the Hermite, Laguerre and Jacobi cases for $\beta$ even according to the
formula
\begin{equation} \label{llave.6}
\rho_{N+1}(y) = (N+1){Q_N \over Z_{N+1}}e^{-\beta V(y)}
f(t_1, \dots, t_\beta) \Big |_{t_1 = \dots = t_\beta = y}
\end{equation}
Let us consider each case in turn, starting with the Jacobi case.

\vspace{.2cm}
\noindent
{\bf Jacobi case}

\vspace{.2cm}
\noindent
Kaneko \cite{kaneko93a} has shown that with
\begin{equation} \label{grifo}
I = [0,1], \quad e^{-\beta V(y)}= y^{\lambda_1} (1 - y)^{\lambda_2},
\qquad t:=(t_1,\ldots,t_m),\quad \la_i=\beta\mu_i/2,\quad (i=1,2)
\end{equation}
$f:=f^{(J)}(\lambda_1,\lambda_2,\beta;t)$ 
as given by (\ref{llave.1}) is the unique solution of each of the p.d.e.'s
\begin{eqnarray}\lefteqn{
t_p(1-t_p){ \partial^2 F \over \partial t_p^2} +
\Big ( c' - {2 \over \beta} (m-1) - (a' + b' + 1 - {2 \over \beta}
(m - 1) ) t_p \Big ) {\partial F \over \partial t_p} - a'b' F}
\nonumber  \\ &&
+ {2 \over \beta} \sum^N_{j = 1 \atop j \ne p}
{1 \over t_p - t_j} \Big ( t_p ( 1 - t_p)
{\partial F \over \partial t_p} - t_j (1 - t_j)
{\partial F \over \partial t_j} \Big ) = 0 \hspace{2cm}
\label{llave.4}
\end{eqnarray}
$(p = 1, \dots, m)$ with
\begin{eqnarray}
a' = -N, \; m = \beta, \;
b' = {2 \over \beta} ( \lambda_1+\lambda_2 +m + 1) + N -1, \; 
c' = {2 \over \beta}(\lambda_1+m) \label{2.2a}
\end{eqnarray} 
Furthermore, Kaneko (see also Yan \cite{yan92b}) has shown  that the solution
of (\ref{llave.4}), normalized to unity at the origin, is 
also given by the generalized hypergeometric function\linebreak
${}_2^{}F_{1}^{(\beta / 2)}(a',b';c';t)$ where
\begin{equation} \label{llave.2}
{}_p^{}F_{r}^{(\beta / 2)}(a_1,\ldots,a_p;b_1,\ldots,b_r;t)
:= \sum_{\kappa} {1 \over |\kappa|!} 
{[a_1]_\kappa^{(\beta / 2)} \cdots [a_p]_\kappa^{(\beta / 2)} \over
[b_1]_\kappa^{(\beta / 2)} \cdots [b_r]_\kappa^{(\beta / 2)}} 
C_\kappa^{(\beta /2)}(t)
\end{equation}
(c.f. (\ref{recordar}) ), so that 
\begin{equation} \label{llave.3}
f^{(J)}(\lambda_1,\lambda_2,\beta;t) = {}_2 F_1^{(\beta/2)}
(-N,\frac{2}{\beta}(\la_1+\la_2+m+1)+N-1;\frac{2}{\beta}
(\la_1+m);t)
\end{equation}
Notice that with $a'=-N$, ${}_2^{}F_{1}^{(\beta / 2)}(a',b';c';t)$
as defined by (\ref{llave.2}) indeed terminates and gives a polynomial.
To see the connection with the generalized Jacobi polynomials, we
note that by summing the p.d.e.'s (\ref{llave.4}) an eigenvalue
equation results. The eigenoperator is precisely the operator
(\ref{ejpoly}) with
$$
N=m,\; a=\frac{2}{\beta}(\la_1+1)-1, \;b=\frac{2}{\beta}(\la_2+1)-1.
$$
Furthermore, from (\ref{llave.2}) and (\ref{llave.3}) 
$f^{(J)}$ has a Jack polynomial
expansion of the form (\ref{gexp}) and from the definition 
(\ref{llave.1}) of $f$, the highest weight monomial in the power
series expansion of $f$ is $m_{(N^m)}$. We thus have
\begin{equation} \label{llave.5}
f^{(J)}(\lambda_1,\lambda_2,\beta;t) = \tilde{G}_{(N^m)}^{(2(\la_1+1)/\beta
-1, 2(\la_2+1)/\beta -1)}(t;\beta/2)
\end{equation}
The tilde here denotes that the normalization in the generalized Jacobi
polynomial is such that it equals unity at the origin.
Comparison of (\ref{llave.3}) and (\ref{llave.5}) gives an 
equality between $\tilde{G}$ and ${}_2 F_1$ therein.

The formula (\ref{llave.6}) for $\rho_{N+1}(y)$ also requires
the value of $Q_N$ and $Z_{N+1}$.
Both quantities are examples of the Selberg integral
\begin{eqnarray}
S_N(\lambda_1,\lambda_2,\lambda) & := &
\left ( \prod_{l=1}^N \int_0^1 dt_l \, t_l^{\lambda_1}(1-t_l)^{\lambda_2}
\right ) \prod_{1 \le j < k \le N} |t_k - t_j|^{2 \lambda}
\nonumber \\
& = &  \prod_{j=0}^{N-1} {\Gamma (\lambda_1 + 1 + j\lambda)
\Gamma (\lambda_2 + 1 + j\lambda)\Gamma(1+(j+1)\lambda) \over
\Gamma (\lambda_1 + \lambda_2 + 2 + (N + j-1)\lambda) \Gamma (1 + \lambda )}.
\end{eqnarray}
We have
\begin{equation} \label{car.2}
Z_{N+1} = S_{N+1}(\beta \mu_1 /2, \beta \mu_2 /2, \beta /2), \quad
Q_N = S_N(\beta \mu_1 /2 + \beta N, \beta \mu_2 /2, \beta / 2).
\end{equation}
Substituting (\ref{llave.5}) and (\ref{car.2}) in (\ref{llave.6}) gives
\begin{eqnarray}
\rho_{N+1}(y) = (N+1) 
 {S_N(\beta \mu_1 /2 + \beta N, \beta \mu_2 /2, \beta /2) \over
 S_{N+1}(\beta \mu_1 /2, \beta \mu_2 /2, \beta /2)}
y^{\beta \mu_1 /2} (1 - y)^{\beta \mu_2 /2} \nonumber \\
\times\tilde{G}_{(N^\beta)}^{(2/\beta +\mu_1-1, 2/\beta +\mu_2-1)}
(t_1, \dots, t_\beta;\beta/2) \Big |_{t_1 = \dots = t_\beta = y}
\label{car.3}
\end{eqnarray}

Our ability to compute the global limit relies on an integral representation
of ${}_2 F_1^{(1/\la)}$ (different of course from
(\ref{llave.1})) and thus, after equating (\ref{llave.3}) 
and (\ref{llave.5}), 
of $\tilde{G}_{(N^m)}$. This integral representation can be derived 
from the integral representation of the generalized hypergeometric function
\cite{yan92b}
\begin{eqnarray}\lefteqn{
{}_2^{}F_1^{(1/\lambda)}(a, \lambda (m-1) + \nu_1 + 1; 2 \lambda 
(m-1) + \nu_1 + \nu_2 + 2; t)} \nonumber \\
& & 
= {1 \over S_m(\nu_1, \nu_2, \lambda)}
\int_{[0,1]^m} dx_1 \dots dx_m \,
{}_1^{}{\cal F}_{0}^{(1/\lambda)} (a;t;x)
D_{\nu_1, \nu_2, \lambda}(x) \label{car.4}
\end{eqnarray}
with ${}_1{\cal F}_{0}^{(1/\lambda)}$ given by (\ref{recordar}), and
\begin{equation}
D_{\nu_1, \nu_2, \lambda}(x)
:= \prod_{j=1}^m x_j^{\nu_1} (1 - x_j)^{\nu_2}
\prod_{1 \le j < k \le m} |x_k - x_j|^{2 \lambda}
\end{equation}
Since  $ \tilde{G}$ in (\ref{car.3}) is equal to ${}_2 F_1^{
(\beta/2)}$ in (\ref{llave.3}) with $m=\beta$, we must set
\begin{equation}
a=-N,\; \lambda = 2/\beta, \; \nu_1 = {4 \over \beta}+\mu_1+\mu_2 + N - 2,
\; \nu_2 = -2  - \mu_2 - N 
\end{equation}
in (\ref{car.4}).

We note that $\nu_2$ is negative so that  (\ref{car.4}) is not defined as
written. However we can readily analytically continue the integral
  (\ref{car.4}) so
that it is valid for  $\nu_2$  negative by following the procedure
detailed in \cite{forr94b}. Thus we deform the contours $[0,1]^m$ to the
contours ${\cal C}^m$, where ${\cal C}$ is any simple closed contour which
starts at the origin and encircles the point $x=1$ (this is first done
under the assumption that $\nu_2$ is not an integer, and $\lambda$
is an integer; it is extended to all $\nu_2$ by analytic
continuation and to all $\lambda$ by noting that the r.h.s.~is analytic
in $\lambda$ when it is defined, while the l.h.s.~is a rational function of
$\lambda$ in the case of interest ($a = -N$)).
Furthermore, we have the formula\cite{kaneko93a} 
\begin{equation}
{}_1^{}{\cal F}_{0}^{(1/\lambda)} (a;t; x)
\Big |_{t_1, \dots, t_\beta=c} = \prod_{l=1}^m (1 - cx_l)^{-a}
\end{equation}
Thus we have
\begin{eqnarray}
\tilde{G}_{(N^\beta)}^{(2/\beta +\mu_1-1, 2/\beta +\mu_2-1)}
(t_1, \dots, t_\beta;\beta/2)
\Big |_{t_1 = \dots = t_\beta = y} = \hspace{6cm}\nonumber\\
{1 \over C} \int_{{\cal C}^\beta} dx_1 \dots dx_\beta
\prod_{l=1}^\beta (1 - yx_l)^N 
x_l^{4/\beta + \mu_1 + \mu_2 + N - 2}
(1 - x_l)^{-2 - \mu_2 - N} \!\!\prod_{1 \le j < k \le \beta}
| x_k - x_j |^{4/\beta} \quad \label{car.6}
\end{eqnarray}
where $C$ is chosen so that at $t=0$, $ \tilde{G}$
is unity. This is our desired integral representation.

\vspace{.2cm}
\noindent
{\bf Laguerre case}

\noindent
Let (\ref{llave.1}) with Laguerre weight $e^{-\beta V(y)}=e^{-\beta y/2}
y^{\beta\mu/2}$ and integration interval $I=[0,\infty)$ be denoted
$f=f^{(L)}(\mu,\beta;t)$. Comparison with the definition of $f^{(J)}$
shows
\begin{equation}
\lim_{L\rightarrow\infty}\left(\frac{1}{L}\right)^{mN}
f^{(J)}(\beta\mu/2,\beta L/2,\beta;t/L) = f^{(L)}(\beta\mu/2,\beta;t)
\end{equation}
Substituting (\ref{llave.3}) and (\ref{llave.5}) for $f^{(J)}$
and using (\ref{llim}) and the fact that\\
$\lim_{b\rightarrow\infty}{}_2 F_1^{(\al)}(a,b;c;x/b) =
{}_1 F_1^{(\al)}(a;c;x)$, we thus have
\begin{equation}
f^{(L)}(\beta\mu/2;t) = \tilde{L}^{\mu-1+2/\beta}_{(N^m)}(t;\beta/2)
={}_1 F_1^{(\beta/2)}(-N;\mu+2;t)
\end{equation}
where $\tilde{L}$ denotes the generalized Laguerre polynomial 
normalized to unity at the origin. (The equality between $f^{(L)}$
and ${}_1 F_1^{(\beta/2)}$ has previously been given in
\cite{forr94b} and the equality between $\tilde{L}^{\mu-1+2/\beta}_{(N^m)}$
and ${}_1 F_1^{(\beta/2)}$ has been noted in \cite{lass91b}.)
Furthermore from the working in \cite{forr94b} we have
\begin{eqnarray}
{}_1 F_1^{(\beta/2)}(-N;\mu+2;t_1,\ldots,t_\beta)
\Big |_{t_1=\dots t_\beta =y} = \nonumber\hspace{6cm}\\
\frac{1}{C'}\int_{({\cal C}')^{\beta}} dx_1\cdots dx_\beta
\prod_{j=1}^{\beta}e^{yx_j} x_j^{-N-3+2/\beta} 
(1-x_j)^{\mu+N+2/\beta -1} \prod_{1\leq j<k\leq\beta}
|x_k-x_j|^{4/\beta} \label{car.7}
\end{eqnarray}

\vspace{.2cm}
\noindent
{\bf Hermite case}

\noindent
Starting with $I$ and $e^{-\beta V(y)}$ given by (\ref{grifo}), the Hermite
case $I = (-\infty, \infty)$ and $e^{-\beta V(y)}= e^{-\beta y^2 /2}$
can be obtained by the change of variables and limiting procedure
\begin{equation} \label{bell.4}
y_j \mapsto {1 \over 2}(1 - {y_j \over L}), \quad
t_j \mapsto {1 \over 2}(1 - {t_j \over L}), \quad
\beta \mu_1 /2 = \beta \mu_2 /2 = \beta L^2 / 2, \quad L \to \infty
\end{equation}
Hence from (\ref{hlim}) and (\ref{llave.5}) with $\la_1=\la_2=
\beta L^2/2$ we see that in the Hermite case, $f^{(H)}$ is
proportional to $H_{(N^m)}(t_1,\ldots,t_m;\beta/2)$ (the fact that
$f^{(H)}$ is an eigenfunction of (\ref{ehpoly}) with eigenvalue $-2N$
was shown in \cite{forr92a}). Thus, from (\ref{p.1}), if we denote
by $\bar{H}_{\ka}$ the generalized Hermite polynomial normalized
so that the coefficient	of the highest weight monomial $m_{\ka}$ is
unity, we have
\begin{subeqnarray}
\rho_{N+1}(x) = (N+1) {Z_N \over Z_{N+1}} e^{-\beta x^2 / 2}
\bar{H}_{(N^\beta)}(t_1, \dots, t_\beta;\beta/2) \Big |_{t_1 =
\dots t_{\beta} = x} \hspace{2cm}\slabel{bell.1} \\
\mbox{where}\hspace*{3cm}
Z_N =  \prod_{l=1}^N \int_{-\infty}^\infty d \lambda_l \,
\exp \left (-{\beta \over 2} \sum_{j=1}^N
\lambda_j^2 \right ) \prod_{1 \le j < k \le N}
|\lambda_k - \lambda_j|^\beta \hspace{2cm}\nonumber \\
 =   {\beta}^{-N/2 - N\beta (N-1)/4}
(2\pi)^{N/2} \prod_{j=0}^{N-1} \frac{\Gamma(1+\beta(j+1)/2)}
{\Gamma(1+\beta/2)} \hspace{2cm}\slabel{bell.2}
\end{subeqnarray}
(compare (\ref{bell.2}) with ${\cal N}_0^{(H)}$ in Proposition 3.7).
To obtain a form of $\bar{H}_{(N^\beta)}$ suitable for asymptotic analysis,
we make use of the integral representation Corollory 3.2 of $H_\kappa$. 
In the case of interest ($\kappa = (N^\beta)$, $t_1 = \dots = t_\beta=x$)
we have
$$
{}_0^{}{\cal F}_{0}^{(2/\beta)} (2y_1, \dots, 2y_\beta; -iz_1,\dots,
-iz_\beta)
\Big |_{z_1 = \dots = z_\beta = x} = \prod_{j=1}^{\beta} e^{-2ixy_j}
$$
and
$$
C_{(N^\beta)}^{(2/\beta)}(iy_1, \dots, iy_\beta)= \prod_{j=1}^{\beta} 
(i y_j)^N
$$
so we can complete the square in the integrand of the formula of
Corollary 3.2 and change variables to obtain
\begin{equation}
\bar{H}_{(N^\beta)} (t_1, \dots, t_\beta;\beta/2)
\Big |_{t_1 =\dots=t_\beta = x} 
= {1 \over V_\beta}  \int_{ {\rr}^\beta} du_1 \dots du_\beta 
\prod_{j=1}^\beta (iu_j + x)^N e^{-u_j^2} \prod_{1 \leq j < k \leq \beta}
|u_k - u_j|^{4/\beta}\label{bell.3}
\end{equation}
where 
\begin{equation}
V_m := {\cal N}_0^{(H)}(N=m,\alpha = \beta/2) \label{bell.3.5}
\end{equation} 
(It is also possible to derive (\ref{bell.3}) by performing the
limiting procedure (\ref{bell.4}) in the integral 
representation (\ref{car.6}).)

Analogous to the situation  
in (\ref{car.6}) and (\ref{car.7}), we note that each integration path along
the real line can be deformed to the path ${\cal C}''$, where
${\cal C}''$ is a simple contour which starts at $-\infty$ and ends at
$\infty$ (this is true for $2/\beta \in\zz_{\ge 0}$ by Cauchy's theorem; it
then follows for all values of $2/\beta$ that the r.h.s.~is defined  by noting
that the r.h.s.~is then analytic in $2/\beta$ while the
l.h.s.~is a rational function in this variable).

\subsection{The global density limit}

Using the integral representations (\ref{car.6}), (\ref{car.7}) 
and (\ref{bell.3}),
the global density limits in (\ref{wig.h}), (\ref{wig.l}) and
(\ref{wig.j}) can be computed for all $\beta$ even. 
The method used in each case is to deform
the contours so that they pass through the saddle points (for each integration
variable there are two saddle points), and to expand the integrand in the
neighbourhood of these points. Due to the similarities of the three
calculations, we will give the details in the Hermite case only.

Changing variables $u_l \mapsto \sqrt{2N} u_l$ in the integral 
representation (\ref{bell.3}), and substituting the result 
in (\ref{bell.1}) gives
\begin{eqnarray}
\rho_{N+1}(\sqrt{2N} x) &=& (N+1){Z_N \over Z_{N+1} V_\beta}
(2N)^{(\beta N + 3 \beta - 2)/2} e^{-\beta N x^2}
\nonumber \\ & & \times
\int_{{\rr}^\beta} du_1 \dots du_\beta 
\prod_{l=1}^\beta e^{-2N u_l^2} (iu_l + x)^N 
\prod_{1 \le j < k \le \beta} |u_k - u_j|^{4/\beta}
\label{bell.5}
\end{eqnarray}
For each integration variable $u_l$ the $N$-dependent terms in the
integrand are
$$
e^{-2N u_l^2} (iu_l + x)^N = e^{-2N u_l^2 + N \log (i u_l + x)} 
$$
The exponent has a stationary point when
\begin{equation}
u_l = u_{\pm} := {ix \over 2} \pm {1 \over 2} (1 - x^2)^{1/2},
\label{bell.6}
\end{equation}
so according to the saddle point method of asymptotic analysis we should
deform each of the contours of integration in (\ref{bell.5}) to pass through
$u_+$ and $u_-$.

With the contours of integration so deformed, we must expand the integrand
in the neighbourhood of the saddle points. Due to the factor
$\prod_{1 \le j < k \le \beta} |u_k - u_j|^{4/\beta}
$ the maximum contribution will be obtained by expanding
$\beta / 2$ integration variables  
($u_1, \dots, u_{\beta /2}$ say) about $u_+$ and the remaining 
$\beta / 2$ integration variables ($u_{\beta /2 +1}, \dots, u_{\beta}$)
about  $u_-$. This specific choice is only one of the $\left (
{\beta \atop \beta / 2} \right )$ equivalent ways of dividing the
integration variables into these two classes, so  after expanding the
variables with the specific choice  we must multiply by the
combinatorial factor. 

In the neighbourhood of the specified points we have 
\begin{eqnarray}\lefteqn{
 e^{-2N u_l^2 + N \log (i u_l + x)}
}\nonumber \\ & & 
 \sim \exp[ -2N u_{\pm}^2 + N \log (i u_{\pm} + x) - {1 \over 2}
( u - u_{\pm})^2 ( 4N - {N \over (i u_{\pm} + x)^2} )]
\label{pali.1}
\end{eqnarray}
where on the r.h.s.~$u_+$ is to be taken for $j=1, \dots, \beta /2$ while
$u_-$ is to be taken for $j= \beta /2+1 , \dots, \beta$. Also
\begin{equation}
\prod_{1 \le j < k \le \beta} |u_k - u_j|^{4/\beta}
\: \sim \: |u_+ - u_-|^{\beta} \prod_{1 \le j < k \le \beta/2} |u_k -
 u_j|^{4/\beta} \prod_{\beta/2+1 \le j < k \le \beta} |u_k - u_j|^{4/\beta}
\label{pali.2}
\end{equation}
Thus after substituting
(\ref{pali.1}) and (\ref{pali.2}) in (\ref{bell.5}) we obtain
\begin{eqnarray}
\rho_{N+1}(\sqrt{2N} x)  \sim \hspace{10cm} \nonumber \\
 N{Z_N \over Z_{N+1} V_\beta}
(2N)^{(\beta N + 3 \beta - 2)/2} e^{-\beta N x^2}
e^{-N \beta (u_+^2 + u_-^2) + (N \beta /2)\log|iu_+ + x|^2}
|u_+ - u_-|^\beta \left (
{\beta \atop \beta / 2} \right ) \nonumber \qquad\\
\times \left|\int_{{\rr}^{\beta/2}} du_1 \dots du_{\beta/2}
\prod_{l=1}^{\beta/2} \exp[-2N u_l^2 (2N - {N \over 2 (iu_+ + x)^2})]
\prod_{1 \le j < k \le \beta/2} |u_k - u_j|^{4/\beta} \right|^2
\qquad\label{pali.3}
\end{eqnarray}

To simplify (\ref{pali.3}) note that a simple change of variables gives that
the last line is equal to
$$
\left( {1 \over | 2N - {N \over 2(iu_+ + x)^2} |} \right)^{\beta - 1}
(V_{\beta / 2})^2,
$$
where $V_{\beta / 2}$ is defined by (\ref{bell.3.5}).
Now suppose $x<1$ so that $u_-^* = - u_+$. Using
(\ref{bell.6}) we then have
$$
 \bigg | 2N - {N \over 2(iu_+ + x)^2} \bigg | = 4N(1 - x^2)^{1/2}, \quad
|u_+ - u_-| = (1 - x^2)^{1/2}, \quad
$$
$$
u_{+}^2 + u_-^2 = {1 \over 2} - x^2, \quad
|iu_+ + x| = {1 \over 2}.
$$
Making these substitutions in  (\ref{pali.3}) shows that
\begin{equation}
\rho_{N+1}(\sqrt{2N} x) \sim (1-x^2)^{1/2} N {Z_N \over Z_{N+1}}
{(V_{\beta / 2})^2 \over V_{\beta}} \left (
{\beta \atop \beta / 2} \right ) e^{-N \beta / 2}
2^{-N \beta / 2 - \beta /2 + 1} N^{\beta (N + 1)/2}
\label{pali.4}
\end{equation}
To simplify the $x$-independent terms in (\ref{pali.4}) we note from the
specific formula (\ref{bell.2}) and Stirling's formula that
\begin{eqnarray} {Z_N \over Z_{N+1}} & = &
\beta^{(1 + \beta N)/2} (2 \pi)^{-1/2} 
{\Gamma (1 + \beta / 2) \over \Gamma (1 + (N + 1) \beta /2)} \nonumber\\
& \sim & {\Gamma (\beta /2 + 1) \over \pi} 2^{\beta N /2 -1/2}
N^{-(\beta / 2)(N+1) -1/2} (\beta /2)^{-(\beta /2)} e^{N \beta /2}
\label{pali.5}
\end{eqnarray}
Also, from (\ref{bell.3.5}) and 
 straightforward manipulation of the explicit formula in
Proposition 3.7 we have 
\begin{equation}
{(V_{\beta / 2})^2 \over V_\beta} = 2^{\beta / 2}
(\beta / 2)^{\beta / 2} {\Gamma (1 + \beta / 2) \over \Gamma (1 + \beta)}
\label{pali.6}
\end{equation}
Substituting (\ref{pali.5}) and (\ref{pali.6}) in (\ref{pali.4}) gives
\begin{equation}
\rho_{N+1}(\sqrt{2N} x) \sim {\sqrt{2N} \over \pi} (1-x^2)^{1/2},
\qquad |x| \le 1
\label{pali.7}
\end{equation}
which is precisely the formula (\ref{wig.h}) for $|x| \le 1$.

For the
intervals $|x| > 1$, instead of repeating the working of the
expansion about the saddle points (which are both pure imaginary in this
case), we note from the result (\ref{pali.7}) that
$$
\int_{-1}^1 \rho(\sqrt{2N} x) \, d(\sqrt{2N} x) \: \sim \:
{2N \over \pi} \int_{-1}^1 (1 - x^2)^{1/2} dx = N
$$
But from the definition of the density it is non-negative and
satisfies the normalization
$$
\int_{-\infty}^\infty \rho(\sqrt{2N} x) d(\sqrt{2N} x) = N.
$$
Thus we must have
$$
{\rho(\sqrt{2N} x) \over \sqrt{2N}} \to 0, \quad {\rm for} \quad |x| >1,
$$
as predicted by (\ref{wig.h}).

\subsection{Initial value problems}

The summations $G^{(H)}(w,z;t)$ in Proposition 3.9 and
$G^{(L)}(x,y;t)$ in Proposition 4.12 are essentially the 
Green functions for the solution of the 
Fokker-Planck equation (\ref{fp-eqn}) with $W$ given by (\ref{hermpot})
and (\ref{lagpot}) respectively. To see this, we first recall that 
$P=\tilde{G}(x^{(0)}|x;\tau)$ is the Green function solution
of (\ref{fp-eqn}) if it is the solution which satisfies the
initial condition 
$$
P(x;\tau)\Big |_{\tau=0} = \prod_{l=1}^N \delta(x_l-x_l^{(0)}),
\hspace{3cm} x_1^{(0)} <\cdots < x_N^{(0)}
$$

By appluing the transformation
(\ref{fp-op}) the Fokker-Planck equation  can be written as   the Schr\"odinger equation
(\ref{schro}), where $t=\tau/i\beta$.
In general the Green function solution of the Schr\"odinger equation,
$G(x^{(0)}|x;t)$ say, may be written in terms of the
eigenvalues and eigenfunctions of $H$. Thus suppose 
$\{\psi_{\ka}\}_{\ka}$ is a complete set of orthogonal eigenfunctions 
of $H$ with corresponding eigenvalues $\{E_{\ka}\}_{\ka}$. Then the
method of separation of variables gives
$$
G(x^{(0)}|x;t) = \sum_{\ka} \frac{\psi_{\ka}(x^{(0)})
\psi_{\ka}(x)}{{\cal N}_{\ka}} \, e^{-it E_{\ka}} 
$$
Thus for the Fokker-Planck equation
\begin{equation}\label{alto.2}
\tilde{G}(x^{(0)}|x;\tau) = e^{\tau E_0 /\beta} {\psi_0(x) \over
\psi_0(x^{(0)})} G(x^{(0)}|x;\tau/i\beta)
\end{equation}

Now, for the Schr\"odinger operators (\ref{herm1}), (\ref{lag1})
\begin{subeqnarray}\slabel{alto.3}
\psi_{\ka}^{(H)}(x) &=& \psi_0^{(H)}(x) H_{\ka}(x/\sqrt{\al};\al)
\hspace{2cm} E^{(H)}_{\ka} = E^{(H)}_0 + \frac{2}{\al}\,|\ka| \\
\psi_{\ka}^{(L)}(x) &=& \psi_0^{(L)}(x) L^{(a'/\al-1/2)}_{\ka}
(x^2/\al;\al)
\qquad E^{(L)}_{\ka} = E^{(L)}_0 + \frac{4}{\al}\,|\ka| 
\slabel{alto.4}
\end{subeqnarray}
where
\begin{eqnarray*}
\psi_0^{(H)}(x) &:=& \prod_{j=1}^N e^{-x_j^2/2\al}\,
\prod_{1\leq j<k\leq N}|x_k-x_j|^{1/\al} \\
\psi_0^{(L)}(x) &:=& \prod_{j=1}^N x_j^{a'/\al}\,e^{-x_j^2/2\al}\,
\prod_{1\leq j<k\leq N}|x^2_k-x^2_j|^{1/\al} 
\end{eqnarray*}
Substituting (\ref{alto.3}) and (\ref{alto.4}) into (\ref{alto.2})
and comparing with the definitions of \linebreak $G^{(H)}(w,z;t)$
and $G^{(L)}(x,y;t)$ shows that
\begin{subeqnarray} \label{alto.7}
\tilde{G}^{(H)}(x^{(0)}|x;\tau) &=& \alpha^{-Nq/2}
\left(\psi_0^{(H)}(x)\right)^2 \,G^{(H)}(
x^{(0)}/\sqrt{\al},x/\sqrt{\al};e^{-\tau}) \quad\slabel{alto.5}\\
\tilde{G}^{(L)}(x^{(0)}|x;\tau) &=& \alpha^{-N(a'/\alpha -1/2 + q)}
\left(\psi_0^{(L)}(x)\right)^2 \,G^{(L)}(
(x^{(0)})^2/\al,x^2/\al;e^{-2\tau}) \Big |_{a=(a'/\al-1/2)} 
\qquad\slabel{alto.6}
\end{subeqnarray}

{}From (\ref{cat.1}) and (\ref{cat.2}) we see from (\ref{alto.7})
that for some initial conditions it is possible to express
$\tilde{G}^{(H)}$ and $\tilde{G}^{(L)}$ in terms of elementary
functions. Thus for $x^{(0)}=c$ (i.e. $x_1^{(0)} = \cdots =
x_N^{(0)}=c$ ) in the Hermite case,
 from (\ref{cat.1}) and (\ref{alto.5}) we have
\begin{subeqnarray}
\tilde{G}^{(H)}(x^{(0)}|x;\tau) \Big|_{x^{(0)}=c} =
\frac{1}{{\cal N}_0^{(H)}} \left( \alpha(1-e^{-2\tau})
\right)^{-Nq/2}  \hspace{4cm}\nonumber\\
\times \exp\left(-\frac{1}{\al(1-e^{-2\tau})}\sum_{j=1}^N
(x_j-e^{-\tau}c)^2 \right) \prod_{1\leq j<k\leq N}
|x_k-x_j|^{2/\al} \hspace{15mm}\\[2mm]
\mbox{ while for $x^{(0)}=0$ in the Laguerre case,
 (\ref{cat.2}) and (\ref{alto.6})
give\hspace*{3.5cm}} \nonumber \\[2mm]
\tilde{G}^{(L)}(x^{(0)}|x;\tau) \Big|_{x^{(0)}=0} =
\frac{1}{{\cal N}_0^{(L)}}\Big |_{a = a'/\alpha - 1/2}
\left( \alpha(1-e^{-2\tau})
\right)^{-N(a'/\al-1/2+q)}  \nonumber\hspace{2.5cm}\\
\times\prod_{j=1}^N x_j^{2a'/\al}\,\exp\left(
-\frac{1}{\al(1-e^{-2\tau})}\sum_{j=1}^N
x_j^2 \right) \prod_{1\leq j<k\leq N} |x^2_k-x^2_j|^{2/\al} 
\hspace{2cm}
\end{subeqnarray}
Now that these explicit solutions have been revealed, they can
be verified independent of the theory of generalized classical
polynomials, by direct substitution into (\ref{fp-eqn}) with
the appropriate $W$.

Another consequence of (\ref{alto.7}) is that it implies
the asymptotic small-$\tau$ behaviour of
$$
{}_0{\cal F}_0^{(\alpha)} \Big ({x^{(0)} \over \tau^{1/2}};
{x \over \tau^{1/2}} \Big )
\quad {\rm and} \quad {}_0{\cal F}_1^{(\alpha)}(a+q; {(x^{(0)})^2 \over
2 \tau} ; {x^2 \over 2 \tau} \Big ).
$$
Thus in general, as $\tau \to 0$ the asymptotic solution of the
Schr\"odinger equation (\ref{schro}) (with $t = \tau / i\beta$) is given by
$$
G(x^{(0)}|x;\tau/i\beta) \: \sim \:
\Big ( { \beta \over 4 \pi \tau } \Big )^{N/2} 
\prod_{j=1}^N e^{-\beta (x_j - x_j^{(0)})^2 / 4\tau}
$$
Substituting this in (\ref{alto.2}), substituting the result 
in (\ref{alto.7}) and using Propositions 3.9 and 4.12 gives
\begin{equation}
{}_0{\cal F}_0^{(\alpha)} \left({x^{(0)} \over  \tau^{1/2}};
{x \over \tau^{1/2}} \right)
\: \sim \: {\pi^{-N/2} 2^{N(N-1)/2\alpha} {\cal N}_0^{(H)} \over
\Big ( \prod_{1 \le j < k \le N} (x_j - x_k) (x_j^{(0)} -
x_k^{(0)})/\tau \Big )^{1/\alpha}}
\prod_{j=1}^N e^{x_j x_j^{(0)} / \tau}
\end{equation}
and
\begin{eqnarray}
{}_0{\cal F}_1^{(\alpha)}\left(a +q; {(x^{(0)})^2 \over
2 \tau} ; {x^2 \over 2 \tau} \right)
\: \sim \:
{\pi^{-N/2} 2^{N(a+1/2)+N(N-1)/\alpha} {\cal N}_0^{(L)}
\over 
\Big ( \prod_{1 \le j < k \le N} (x_j^2 - x_k^2) ((x_j^{(0)})^2 -
(x_k^{(0)})^2)/ \tau  \Big )^{1/\alpha}} \nonumber\\
\times\;\prod_{j=1}^N (x_j x_j^{(0)}/ \tau)^{-(a+1/2)} 
\prod_{j=1}^N e^{x_j x_j^{(0)} /  \tau} \qquad
\end{eqnarray}
where it is assumed $x_1 < \cdots < x_N$ and $x_1^{(0)} < \cdots <x_N^{(0)}$.
In the case $\alpha = 2$ these asymptotic formulas are known in the mathematical
statistics literature (see e.g.~\cite{muir78a})

\setcounter{equation}{0}
\section{A brief literature survey}

To our knowledge, the generalized classical polynomials were first
introduced by Herz \cite{herz55} in the case $\alpha = 2$ via
integral formulas over measures associated with spaces of orthogonal
matrices (however it should be noted that what Herz calls generalized
Hermite polynomials do not correspond to the generalized Hermite
polynomial we have considered). Constantine and Muirhead extended
the work of Herz on the generalized Laguerre polynomials in the case
$\alpha = 2$, and derived the formulas (\ref{l-gen3}) 
\cite[ex.~7.19]{muir82},
(\ref{l-gen4}) \cite[Thm.~7.6.3]{muir82}, 
(\ref{off.1}) \cite[ex.~7.20]{muir82},
(\ref{off.2}) \cite[eq.~7.6(4)]{muir82}, 
(\ref{tenga}) \cite[Thm.~7.6.5]{muir82},
(\ref{tdd}) \cite[Thm.~7.6.4]{muir82} and 
Proposition 4.12 \cite[ex.~7.21]{muir82}
in that case (in comparing formulas it should be noted that Muirhead
adopts the normalization $c_{\kappa \kappa} = (-1)^{|\kappa|}/
C_\kappa^{(\alpha)}(1^N)$ which is $|\kappa|!$ times the normalization
we have used). For general $\alpha$ and $N=2$ Yan \cite{yan92a} derived
(\ref{l-gen4}) \cite[eq.~(5.6)]{yan92a}
(\ref{off.2}) \cite[last eqn.~p. 251]{yan92a}, 
(\ref{tenga}) \cite[eq.~(5.11)]{yan92a} and
(\ref{td.2}) \cite[eq.~(5.13)]{yan92a} (Yan uses the same normalization as
Muirhead). For general $\alpha$ Lassalle \cite{lass96a} has reported
the results (\ref{offf}) and (\ref{tenga}), 
and simultaneous to our investigations has obtained the 
results (\ref{l-gen3}) and 
(\ref{moops.1}) - (\ref{moops.2}) (Lassalle uses the
normalization $L_\kappa^a(0;\alpha) = 1$). For values of $\alpha$
corresponding to Jordan algebras, generalized Laguerre polynomials
have been studied by Dib \cite{dib90a}.

In an unpublished handwritten manuscript Macdonald \cite{macunp1} has
derived some properties of the generalized classical polynomials. His
results for the generalized Laguerre polynomials, which overlap with the
same equations of ours as does the work of Yan, are typically proved for
$\alpha = 1/2,1$ and 2, and are conjectured to remain valid for general
$\alpha$. The validity of a number of the results in \cite{macunp1}
for general $\alpha$ rely on a conjecture for the so called generalized
Laplace transform of the Jack polynomial:
\begin{eqnarray}
\int_{[0,\infty)^N} {}_0^{}{\cal F}_0^{(\alpha)}(-x;y)
C_\kappa^{(\alpha)}(x) \prod_{j=1}^N x_j^a \prod_{1 \le j < k \le N}
|x_k - x_j|^{2/\alpha} \, dx_1 \dots dx_N  \nonumber\\
= [a+q]_\kappa^{(\alpha)} C_\kappa^{(\alpha)}(1^N)
\prod_{j=1}^N y_l^{-(a+q)} C_\kappa^{(\alpha)}({1 \over y})
\label{f1}
\end{eqnarray}
This conjecture can be proved using results contained herein. First we
calculate the generalized Laplace transform of the generalized Laguerre
polynomial:
\begin{eqnarray}
\int_{[0,\infty)^N} {}_0^{}{\cal F}_0^{(\alpha)}(-x;y)
L_\sigma^a(x;\alpha) \prod_{j=1}^N  x_j^a \prod_{1 \le j < k \le N}
|x_k - x_j|^{2/\alpha} dx_1 \dots dx_N \nonumber\\
= {\cal N}_\sigma^{(L)} \prod_{j=1}^N y_l^{-(a+q)} C_\sigma^{(\alpha)}(1-
{1 \over y}),
\end{eqnarray}
which follows from the first equation of the proof of 
Proposition 4.10 after noting from (\ref{id.7}) that
$$
  \prod_{j=1}^N e^{-x_j}\,  {}_0^{}{\cal F}_0^{(\alpha)}(-x;
{z \over 1 - z}) = {}_0^{}{\cal F}_0^{(\alpha)}(-x;
{1 \over 1 - z})
$$
and writing $1/(1-z) =: y$. We now use (\ref{off.3}) and multiply (\ref{f1})
by a suitable $\sigma$--dependent factor so that after summing over
$\sigma$ we can replace $L_\sigma^a(x;\alpha)$ on the l.h.s.~by
$C_\kappa^{(\alpha)}(x)$. On the r.h.s.~we then have
\begin{equation}
\prod_{j=1}^N y_l^{-(a+q)} [a+q]_\kappa^{(\alpha)}
C_\kappa^{(\alpha)}(1^N)
\sum_{\sigma \subseteq \kappa} \left ( {\kappa \atop \sigma} \right )
{(-1)^{|\sigma|} {\cal N}_\sigma^{(L)}  C_\sigma(1-
{1 \over y}) \over [a+q]_\sigma^{(\alpha)}}
\label{f2}
\end{equation}
Substituting the value of $ {\cal N}_\sigma^{(L)}$ from (\ref{tenga}) 
and using (\ref{gbin}) to compute the sum gives (\ref{f1}) as required. 

The generalized Hermite polynomials of the type considered in this paper
appear to have been first considered by James \cite{james75a}
in the case $\alpha = 2$. Subsequently, for general $\alpha$ Lassalle
\cite{lass91a} noted the orthogonality with respect to the measure
(\ref{ih}), the normalization of Proposition 3.7 and the property of
Proposition 3.3. Furthermore, in handwritten notes Lassalle \cite{lass96a} has
established Proposition 3.1 and has stated Corollaries 3.1 and 3.2,
(\ref{china.1}), (\ref{china.2}) and (\ref{china.7}). 
Also given in the notes is an explicit
formula for the coefficients $c_{\mu \kappa}^{(H)}$ in (\ref{hexp}).
Macdonald \cite{macunp1} has also
considered properties of the generalized Hermite polynomials in the
form of conjectures based on derivations in the cases $\alpha = 1/2$,
 $1$ and $2$.
He has obtained the normalization of Proposition 3.7, the property of
Proposition 3.3 and the integration formula of Proposition 3.8 and the
generating function of Proposition 3.1.

M.~Lassalle has pointed out to us that the exponential operator formulas
(\ref{china.1}) and (\ref{moops.1}) imply an intimate connection
between the theory of generalized Hermite and Laguerre polynomials
and theory developed by Dunkl \cite{dunkl91,dunkl93}.
Inspection of these works show that this is indeed so. The Hermite case
is the most straightforward, which in the language of  \cite{dunkl91,dunkl93}
corresponds to the root system $A_N$. Dunkl introduces the operators
$$
T_i := {\partial \over \partial x_i} + {1 \over \alpha}
\sum_{j=1 \atop j \ne i}^N { 1 - M_{ij} \over x_i - x_j }
$$
where $M_{ij}$ is the operator which exchanges coordinates $x_i$ and
$x_j$. When acting on functions symmetric in $x_1, \dots, x_N$ these
operators are related to $D_0$ (recall (\ref{defs.1})) by
$$
D_0 = \sum_{i=1}^N T_i^2.
$$
Also introduced is the pairing $[p,q]_{H}$. For polynomials
$p$ and $q$ homogeneous of the same degree
\begin{equation}
[p,q]_{H} := p(T^x) q(x), \label{dun.1}
\end{equation}
where $p(T^x)$ means that each variable $x_i$ in $p$ is replaced by $T_i$
(the ordering within the monomials does not matter since the operators
$\{T_i\}$ commute), while if the degrees differ
$$
[p,q]_{H} := 0.
$$
This pairing is intimately related to the exponential operator in 
(\ref{china.1}). Thus, as noted in \cite{lass96a}, it follows from
\cite[Thm.~3.10]{dunkl91} that for homogeneous symmetric polynomials
$p$ and $q$
$$
[p,q]_{H} = {1 \over {\cal N}_0^{(H)} } \int_{(-\infty, \infty)^N}
\Big ( e^{-D_0/4} p \Big ) \Big ( e^{-D_0/4} q \Big ) \, d \mu^{(H)}(x).
$$
{}From (\ref{china.1}) and the orthogonality of $\{H_\kappa\}$ with respect to
the inner product (\ref{ih}) we immediately have the result that the Jack
polynomials are  orthogonal with respect to the pairing (\ref{dun.1}):
$$
[J_\kappa^{(\alpha)}, J_\mu^{(\alpha)}]_{H} = (2 \alpha)^{-|\kappa|}
j_\kappa J_\kappa^{(\alpha)}(1^N) \delta_{\kappa, \mu}
$$ 
where we have used Proposition 3.7 and (\ref{farfel})

Dunkl also introduces a kernel $K(x,y)$, which for the root system
$A_N$ and $p$ a symmetric homogeneous polynomial has the property
\cite[Prop.~2.1]{dunkl93}
$$
p(y) = {e^{-p_2(y)} \over {\cal N}_0^{(H)} } \int_{(-\infty, \infty)^N}
\Big ( e^{-D_0/4} p \Big ) K(x,y) \, d \mu^{(H)}(x)
$$
(we have changed variables $x \mapsto \sqrt{2}x$, $y \mapsto \sqrt{2}y$
and replaced $K( \sqrt{2}x,  \sqrt{2}y)$ by $K(x,y)$). Comparison
with Corollary 3.1 (after substituting for $H_\kappa$ using (\ref{china.1}))
gives the explicit formula
$$
K(x,y) = {}_0^{}{\cal F}_0^{(\alpha)}(2x;y).
$$

In the Laguerre case there are analogous connections with the work of
Dunkl, with the underlying root
system now being $B_N$. The operators $T_i$ are now given by
(see e.g.~\cite{hikami96})
\begin{equation}
T_i = {\partial \over \partial x_i} + {1 \over \alpha}
\sum_{j=1 \atop j \ne i}^N \Big (
{1 - M_{ij} \over x_i - x_j} + { 1 - S_iS_j M_{ij} \over
x_i + x_j} \Big ) + {a + 1/2 \over x_i} (1 - S_i)
\label{dop}
\end{equation}
where the action of the operator $S_i$ is to replace the variable $x_i$
by $-x_i$. When acting on a function $f$ symmetric and even in $x_1, \dots, x_N$ these operators are such that 
$$
\sum_{i=1}^N T_i^2 \Big |_{x_i^2 = u_i} = 
4\Big ( D_1^{(u)} + (a+1)E_0^{(u)} \Big ).
$$
Using (\ref{dop}) and setting $P(x) := p(x^2)$ and $Q(x) = q(x^2)$,
\cite[Thm.~3.10]{dunkl91} gives
$$
[P,Q]_{L} = {1 \over {\cal N}_0^{(L)}} 
\int_{[0,\infty)^N}
\Big ( e^{-(D_1+(a+1)E_0)} p \Big ) \Big ( e^{-(D_1+(a+1)E_0)} q \Big )  
d\mu^{(L)}(x),
$$
where $[\,,\,]_{L}$ is defined by (\ref{dun.1}) with $T_i$ specified by
(\ref{dop}). Orthogonality of $\{L_\kappa\}$ with respect to (2.21)
and use of (\ref{moops.1}) then gives
$$
[J_\kappa^{(\alpha)}, J_\mu^{(\alpha)}]_L =
\alpha^{-|\kappa|} j_\kappa [a+q]_\kappa^{(\alpha)} J_\kappa^{(\alpha)}(1^N)
\delta_{\kappa, \mu}.
$$
Also, (\ref{td.1}) with the substitution of (\ref{moops.1}) and the change of
variables $x,z_a \mapsto x^2,z_a^2$
gives the kernel $K(x,,y)$ in the
$B_N$ case as 
${}_0^{}{\cal F}^{(\alpha)}_1(a+q;x^2;-y^2)$.

In the context of the Calogero-Sutherland model the expansion of the
generalized Hermite polynomials in terms of monomial symmetric
functions has been considered by Ujino and Wadati \cite{uji95a}, and a
Rodrigues--type formula has been obtained \cite{uji96a}, analogous 
to that recently given by Lapointe and Vinet \cite{vinet96a} for
Jack polynomials.
Also Polychronakos \cite{poly96a} has recently considered the
monomial expansion of $H_{(1^k)}(x;\alpha)$ and given its normalization.
To our knowledge there have been no previous discussions of the
generalized Laguerre polynomials in the context of the
Calogero-Sutherland model. 

Regarding the global limit of the density computed in Section 5, we
know of no other works which consider this limit directly. However,
using techniques from potential theory Johansson \cite{Joh1} has recently
proved that, in the Hermite case, for all $\beta \ge 0$
$$
\lim_{N \to \infty} \sqrt{2 \over N} \int_{-\infty}^\infty
f(\sqrt{2N} x) \rho(\sqrt{2N}x) \, dx = {2 \over \pi}
\int_{-1}^1 f(x) \sqrt{1 - x^2} \, dx.
$$
(for $\beta = 2$ this result was first given in \cite{BPS} using a
mean-field approach) for any continuous, bounded $f$. The analogous result
in the Jacobi case has also been obtained by  Johansson \cite{Joh2}.
These results establish that the smoothed density is that predicted by
electrostatics, while our result establishes pointwise convergence
to the electrostatic prediction.

\vspace{3mm}\noindent
{\Large\bf Acknowledgements}\\[2mm]
We are particularly thankful to M. Lassalle for providing us with
\cite{lass96a}, encouraging our research and for advice. We also thank
N. Obata for a useful remark. THB would like to thank Prof. T. Miwa for
hospitality at RIMS where part of this work was carried out.
The financial support of the ARC is acknowledged.

\vspace{1cm}
\setcounter{section}{1}
\setcounter{equation}{0}
\renewcommand{\thesection}{\Alph{section}}

\noindent{\Large\bf Appendix}

\vspace{2mm}\noindent
In this appendix, the equations (\ref{goo.1}), (\ref{goo.2}) and
(\ref{tulum}) are derived as special cases of a p.d.e. satisfied
by ${}_2{\cal F}_1^{(\al)}(a,b;c;x;y)$ (recall (\ref{recordar})).

\vspace{2mm}\noindent
{\bf Proposition A.1} \\[2mm]
{\it Let ${}_2{\cal F}_1^{(\al)}(a,b;c;x;y)$ be defined by (\ref{recordar}).
This function satisfies the p.d.e.
\begin{equation} \label{cofre}
D_1^{(x)}\,F + \left( c-\frac{N-1}{\al} \right)E_0^{(x)}\,F
-\left( a+b - \frac{N-1}{\al} \right)E_2^{(y)}\,F 
-\eta_2^{(y)}\,F = abp_1(y)\,F ,
\end{equation}
where $D_k$, $E_k$ are defined by (\ref{defs.1}) and 
$\eta_2:=\frac{1}{2}[D_2,E_2]$, and is in fact the unique solution
of the equation of the form
\begin{equation} \label{palta}
F(x,y) = \sum_{\ka} A_{\ka}\:\frac{C_{\ka}^{(\al)}(x)
C_{\ka}^{(\al)}(y)}{C_{\ka}^{(\al)}(1^N)},
\hspace{2cm} A_0=1
\end{equation}
}

\vspace{2mm}\noindent
{\it Proof}\quad We follow the method of Constantine and Muirhead%
\cite{muir82} in the case $\al=2$. With $F$ given by (\ref{palta}),
from (\ref{actions}) and (\ref{id.oo}) we have
\begin{subeqnarray}\label{dedos}
D_1^{(x)}\,F &=& \sum_{\ka}\sum_{i=1}^N \bin{\ka^{(i)}}{\ka}
\left( \ka_i +\frac{N-i}{\al} \right)
\frac{C^{(\al)}(x)}{C^{(\al)}(1^N)}\,C^{(\al)}_{\ka^{(i)}}(y)\,
A_{\ka^{(i)}} \\
E_0^{(x)}\,F &=& \sum_{\ka}\sum_{i=1}^N \bin{\ka^{(i)}}{\ka}
\frac{C^{(\al)}(x)}{C^{(\al)}(1^N)}\,C^{(\al)}_{\ka^{(i)}}(y)\,
A_{\ka^{(i)}} \\
E_2^{(y)}\,F &=& \frac{1}{1+|\ka|}\sum_{\ka}\sum_{i=1}^N 
\bin{\ka^{(i)}}{\ka} \left( \ka_i -\frac{i-1}{\al} \right)
\frac{C^{(\al)}(x)}{C^{(\al)}(1^N)}\,C^{(\al)}_{\ka^{(i)}}(y)\,A_{\ka} \\
\eta_2^{(y)}\,F &=& \!\!\frac{1}{1+|\ka|}\sum_{\ka}\sum_{i=1}^N 
\bin{\ka^{(i)}}{\ka} \left( \ka_i -\frac{i-1}{\al} \right)
\left( \ka_i -\frac{i-N}{\al} \right)
\frac{C^{(\al)}(x)}{C^{(\al)}(1^N)}\,C^{(\al)}_{\ka^{(i)}}(y)\,A_{\ka} 
\hspace{1.2cm}\\
p_1(y)\,F &=& \frac{1}{1+|\ka|}\sum_{\ka}\sum_{i=1}^N 
\bin{\ka^{(i)}}{\ka} \frac{C^{(\al)}(x)}{C^{(\al)}(1^N)}\,
C^{(\al)}_{\ka^{(i)}}(y)\,A_{\ka} 
\end{subeqnarray}
Substituting (\ref{dedos}) in (\ref{cofre}), equating coefficients of
$C^{(\al)}_{\ka}(x)/C^{(\al)}_{\ka}(1^N)$ and then equating
coefficients of $C^{(\al)}_{\ka^{(i)}}(y)\bin{\ka^{(i)}}{\ka}$
gives
$$
\left( c+\ka_i -\frac{i-1}{\al} \right)\,A_{\ka^{(i)}} =
\frac{1}{1+|\ka|}\left(a+\ka_i -\frac{i-1}{\al} \right)
\left(b+\ka_i -\frac{i-1}{\al} \right)\,A_{\ka}
$$
This is a first order difference equation and so has a unique
solution once the initial condition $(A_0=1)$ is specified. It
is straightforward to verify that the solution is
$$
A_{\ka} = \frac{1}{|\ka|!}\,\frac{[a]^{(\al)}_{\ka}
[b]^{(\al)}_{\ka}}{[c]^{(\al)}_{\ka}}
$$

The equation (\ref{tulum}) for ${}_1{\cal F}_1^{(\al)}(a;c;x;y)$
follows from (\ref{cofre}) by changing variables $y\rightarrow y/b$
and then taking $b\rightarrow\infty$. The equation (\ref{goo.2})
for ${}_0{\cal F}_0^{(\al)}(x;y)$ follows from that for
${}_1{\cal F}_1^{(\al)}(a;c;x;y)$ by setting $a=c=(N-1)/\al$, while
the equation (\ref{goo.1}) follows from (\ref{tulum}) (with $a$ and
$c$ interchanged) for ${}_1{\cal F}_1^{(\al)}(a;c;x;y)$ by changing
variables $y\rightarrow y/a$ and taking $a\rightarrow\infty$.

\bibliographystyle{plain}

\end{document}